\documentclass[reprint,amsfonts,amssymb, amsmath,   nofootinbib,pra,aps,superscriptaddress, twocolumn,longbibliography]{revtex4-2}

\usepackage{amsmath}
\usepackage{amsthm}
\usepackage{amssymb}
\usepackage{physics}
\usepackage{braket}
\usepackage{xcolor}
\usepackage{url}
\usepackage[colorlinks=true,citecolor=blue,linkcolor=magenta,linktocpage=true]{hyperref}
\usepackage{mathtools}
\usepackage{bbm}
\usepackage[normalem]{ulem}
\usepackage{cleveref}
\usepackage{enumitem}
\usepackage{graphicx}

\newcommand{\para}[1]{\medskip \noindent \textbf{#1 -}}

\newcommand{\mc}[1]{\mathcal{#1}}
\newcommand{\mbb}[1]{\mathbb{#1}}

\newcommand{\dket}[1]{|#1\rangle \! \rangle}
\newcommand{\dbra}[1]{\langle \! \langle #1|}
\newcommand{\dketsp}[1]{|#1\,\rangle \! \rangle}
\newcommand{\hilb}{\mathcal{H}}
\newcommand{\hilbop}{\mathcal{B}}

\newcommand{\homm}{{\rm Hom}}
\newcommand{\aff}{{\rm Aff}}

\newcommand{\homvw}{\homm(V, W)}

\newcommand{\affvw}{\aff(V, W)}

\newtheorem{definition}{Definition}

\renewcommand\set[1]{\{ #1 \}}

\begin{document}
\title{Quantum machine learning models for graphs}
\author{Fr\'{e}d\'{e}ric Sauvage}
\affiliation{Quantinuum, Partnership House, Carlisle Place, London SW1P 1BX, United Kingdom}
\author{Pranav Kalidindi}
\affiliation{Quantinuum, Partnership House, Carlisle Place, London SW1P 1BX, United Kingdom}
\author{Frederic Rapp}
\affiliation{Quantinuum, Partnership House, Carlisle Place, London SW1P 1BX, United Kingdom}
\author{Mart\'{i}n Larocca}
\affiliation{Theoretical Division, Los Alamos National Laboratory, Los Alamos, New Mexico 87545, USA}

\begin{abstract}
Geometric Machine Learning (GML) successes have been achieved through the thorough study and design of new equivariant neural networks. In comparison, geometric quantum machine learning (GQML) models lack such a detailed understanding and, despite already several proposals, a unifying perspective on their design remains elusive. 
In this work, we focus on GQML models for graph problems that showcase a lot of structure and still remain frontier in machine learning.
For the case when $n$--node graphs are encoded in $n$--qubit states, we provide a comprehensive characterization of their constituents. Taken together, these furnish us with a toolbox for the design of  quantum graph models, and we further probe its benefits including the natural integration with classical models, generalization of known GQML models (sometimes extending their expressivity at virtually no cost), and straightforward classical pre-training strategies. The latter two features are demonstrated in dedicated numerical experiments.
\end{abstract}
\maketitle

\section{Introduction}
Despite the potential of quantum machine learning (QML)~\cite{biamonte2017quantum,benedetti2019parameterized,cerezo2022challenges} to widen the reach of quantum technology, scalability of QML models, and thus their applicability to problems of importance, is known to be limited by issues of trainability~\cite{mcclean2018barren,anschuetz2022quantum,larocca2025barren}.
These can be tempered by designing models that are tailored to the problem at hand. In the presence of symmetries in the underlying problem, a common situation in physical and synthetic problems, one desires to constrain the model to \emph{respect the same symmetries}. This is the foundation of Geometric Quantum Machine Learning (GQML) ~\cite{larocca2022group,meyer2023exploiting,zheng2023speeding,west2024provably,le2025symmetry}, which seeks to leverage the insights and emulate the success of classical geometric machine learning (GML)~\cite{cohen2016group,kondor2018generalization,bronstein2021geometric}. 

Rather than a monolithic framework, under the requirement of respecting the symmetries, achieved through \emph{equivariance}~\cite{cohen2016group,kondor2018generalization,bronstein2021geometric,nguyen2024theory}, lies a lot of freedom. 
GML’s successes have been achieved by the comprehensive study and the design of new equivariant layers~\cite{cohen2017steerable,maron2018invariant,finzi2021practical,bouritsas2022improving}. In comparison, the design of GQML models remains largely unexplored.
In this work, we wish to address this deficiency by focusing on graph problems.
Despite their wide range of applications, graph problems remain a frontier in ML, with well documented limitations of graph neural networks (GNNs)~\cite{xu2018powerful,morris2019weisfeiler,alon2020bottleneck,zhu2020beyond,puny2023equivariant}. Furthermore, these showcase many structures that can be exploited when designing ML models. As such, graph problems provide an ideal playground for the study and applications of GQML principles.

\begin{figure*}
\includegraphics[width=0.95\textwidth]{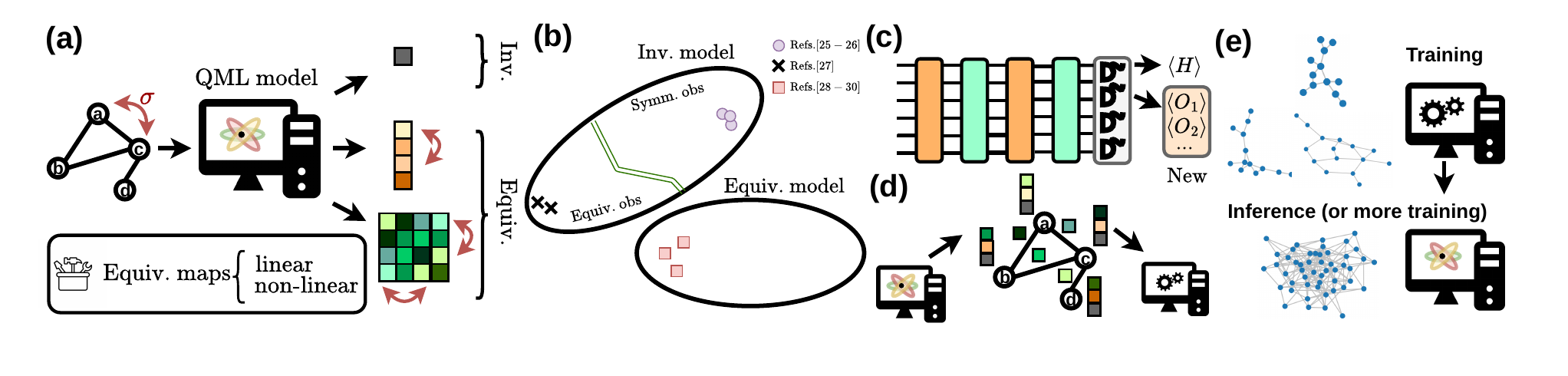}
\vspace{-10pt}\caption{ \textbf{(a)} Toolbox for the construction of GQML models on graphs including the characterization of the relevant equivariant linear and non-linear maps. These are used for the encoding of $n$-node graphs into $n$-qubit quantum states and to identify measurements leading to invariant or equivariant models.  This toolbox allows us to recover known models, identify new ones \textbf{(b)}, and even render existing models more expressive at no cost \textbf{(c)}.
 It also ensures integration with classical models through feature augmentation \textbf{(d)}. The models obtained lend themselves to classical pre-training strategies that can reduce the training effort that needs to be performed on a quantum device and mitigate issues of concentrations.
}
\label{fig:summary}
\end{figure*}

Already several works have explored problems on graphs through GQML, or related quantum heuristics~\cite{mills2019quantum,schatzki2024theoretical,szegedy2019qaoa,mernyei2022equivariant,skolik2023equivariant,albrecht2023quantum,thabet2023enhancing}. 
Still, it remains unclear how the different models proposed to date relate to each other, what are their underlying design principles, and ultimately how they can be generalized. 
As sketched in Fig.~\ref{fig:summary}, the main contribution of this work is to provide a toolbox for the design of GQML models on graphs.
This toolbox relies on a comprehensive identification of the relevant linear and non-linear equivariant maps. 
Equipped with such characterization, we recover known models and discover new ones.
As we further show, benefits of this toolbox include: the natural integration with classical models (while still respecting symmetry principles), extending the expressiveness of known models (sometimes at virtually no cost), and the development of classical pre-training strategies that can reduce quantum computing resource requirements during training and improve on issues of concentration of the output of the models, known as barren plateaus~\cite{mcclean2018barren,larocca2025barren}.

In Sec.~\ref{sec:background}, we provide background for GQML on graphs.
In Sec.~\ref{sec:toolbox}, we detail an exhaustive characterization of the relevant equivariant maps.
These maps constitute the building blocks of a GQML toolbox for graphs. As detailed in Sec.~\ref{sec:composition}, this toolbox recovers and generalizes existing architectures, while also enabling natural interfacing with GNNs.
We further demonstrate its appeal in minimal numerical studies: 
our precise understanding of the equivariant maps allows us to extend the expressivity of current models without incurring any additional circuit or sampling complexity overhead. 
Additionally, we show how the constructed models lend themselves to natural pretraining strategies that can mitigate issues of barren plateaus.
End-to-end numerical studies and implementations are left to future work.

\section{Background}\label{sec:background}
\subsection{Geometric machine learning}
Consider a supervised learning problem: Given access to a data set $\mc{D} = \{ x_i , y_i\}$ consisting of $N$ pairs of input $x_i$ and labels $y_i=f(x_i)$, one aims at \textit{learning} the unknown labeling function $f:\mc{X} \rightarrow \mc{Y}$. 
To do so, a parameterized \textit{model} $h_{\theta}$, with a set of parameters $\theta$, 
is trained to minimize deviations between labels $y_i$ and predictions $h_{\theta}(x_i)$ on the dataset.
General parameterized models are more easily thought about in terms of a composition
\begin{equation}\label{eq:ml_model}
    h_{\theta}(x) = \Lambda_{\theta}^M \circ \Lambda^2_{\theta} \circ \hdots \circ \Lambda^1_{\theta}(x)
\end{equation}
of maps $\Lambda_{\theta}^m: \mc{X}^{(m)} \rightarrow \mc{X}^{(m+1)}$ such that $\mc{X}^{1}=\mc{X}$ and $\mc{X}^{M+1}=\mc{Y}$. In ML, typical inputs and outputs are tensors of real-value features, and the maps are parameterized linear, or more generally affine, 
operations and fixed non-linear ones. 

In addition to the dataset, one often has additional knowledge regarding the function to be learnt. This typically comes as some form of equivariance of $f$ under transformations of its inputs.
\begin{definition}[Equivariance]\label{def:equiv}
Let $G$ be a group of symmetries with action (denoted with a $\cdot$) on $\mc{X}$ and $\mc{Y}$. 
We say that a map $f: \mc{X}\mapsto \mc{Y}$ is equivariant w.r.t. $G$ whenever 
\begin{equation}\label{eq:equiv}
f(g \cdot x) = g \cdot f(x)    
\end{equation}
for any $g\in G$ and any $x \in \mc{X}$. The map is said to be invariant whenever $f( g \cdot x) = f(x)$.
\end{definition}
We stress that the action on $\mc{X}$ and $\mc{Y}$ of the group elements may not be the same. The invariance case corresponds to the action on any $y \in \mc{Y}$ is \emph{trivial} with $g \cdot y = y$. Definition~\ref{def:equiv} is extended to \emph{parameterized functions} $f_{\theta}$, such as our models or its components~\eqref{eq:ml_model}, when  condition~\eqref{eq:equiv} hold for arbitrary values of the parameters $\theta$.

When a problem admits symmetries ($f$ is equivariant or invariant), it is sensible to enforce the model $h_{\theta}$ to respect the same symmetries. This has the benefits of reducing the search space to a subset of relevant functions and ultimately easing the training and generalization of the models~\cite{cohen2016group,bronstein2021geometric}.
For instance, convolutional NNs are often regarded as a major milestone in training ML models on a truly large scale~\cite{krizhevsky2012imagenet}.
This is particularly desirable for QML models often plagued by issues of trainability~\cite{cerezo2022challenges}.

Designing equivariant (or invariant) models is achieved constructively by requiring the intermediate maps~\eqref{eq:ml_model} to be equivariant and the final one to be equivariant (or invariant). This principle forms the basis of GML, a field that has thrived in identifying symmetries of interest and designing layers of neural networks (NNs), equivariant and invariant under relevant symmetries~\cite{cohen2016group,kondor2018generalization,maron2018invariant,morris2019weisfeiler,cohen2017steerable,finzi2021practical}. 
We stress that under the GML program, there is a lot of freedom, including (i) the choice of the intermediary domains $\mc{X}^{m}$, together with (ii) the way equivariant maps between different domains are parameterized.

\subsection{Geometric quantum machine learning}
In Quantum Machine Learning (QML), some of the maps composing the model~\eqref{eq:ml_model} are realized through quantum operations. These include the encoding  of classical data into quantum states, quantum evolution, and the decoding of quantum states back to classical data through measurements.
Combining these, a template QML model is defined as follows.
\begin{definition}[Template QML model]\label{def:base_exp_model}
A prototypical QML model $q: \mc{X}\mapsto \mathbb{R}$, consists in a parameterized map encoding data into a n-qubit state $x\mapsto \rho_\alpha(x)$, a parameterized circuit $U_{\theta}$ and an observable $O$. Denoting $O_{\theta} = U_{\theta}^{\dagger}O U_{\theta}$, the  outputs of the model is given by:
\begin{equation}\label{eq:min_qml}
    q_{\alpha, \theta}(x) := {\rm Tr } \left[ O_\theta  \rho_{\alpha}(x) \right].
\end{equation}
\end{definition}

While Eq.~\eqref{eq:min_qml} is often used to approximate the function $f$ to be learnt, it seems counter-productive to expect such models to solve a task end-to-end. After all, classical ML is already remarkably accurate in many tasks. Rather, it seems more sensible to envision the quantum component, such as Eq.~\eqref{eq:min_qml}, to be part of a larger, and mostly classical, model. In the context of GML, this requires the ability to naturally interface quantum and classical components, while still satisfying equivariance.

The concepts of equivariance and invariance equally applies to classical or quantum models;
We talk about GQML whenever a QML model satisfies the precepts of GML, as per Definition~\ref{def:equiv}. Core to the construction of GQML models is the concept of \emph{symmetric} and \emph{equivariant} observables and maps that are defined in Sec.~\ref{sec:symm_inv_operators}.
Before that, we introduce some preliminaries in Sec.~\ref{sec:notations}.

\subsubsection{Preliminaries}\label{sec:notations}
Let $\hilb=(\mathbb{C}^2)^{\otimes n}$ be the state space of $n$ qubits and let $\tilde{\hilbop}$ be the  space of linear operators acting on $\hilb$. Among them, we will mostly restrict our attention to the subspace of observables (i.e. Hermitian operators) $\hilbop \subset \tilde{\hilbop}$.
For generic vector spaces $V$ and $W$, let $\homvw$ denote the linear maps from $V$ to $W$, and $\affvw$ the affine maps from $V$ to $W$. Let ${\rm End}(V)$ denote the linear maps from $V$ to $V$ itself, such that $\tilde{\hilbop}={\rm End}(\hilb)$. Let ${\rm GL}(V)$ denote  the group of invertible matrices on $V$.

Given a group $G$ of symmetries with action on a vector space $V$ we denote let $R_V(g)$ be the representation of $g \in G$ on $V$. 
Formally a representation is a group homomorphism: a map $G \mapsto {\rm GL}(V)$ between the groups $G$ and ${\rm GL}(V)$ that preserves the group structure $R_V(gh)=R_V(g)R_V(h)$ for any $g$ and $h\in G$ (it essentially embeds $G$ into ${\rm GL}(V)$, not necessarily in an injective fashion).
Action on $V$ can thus be expressed through $g \cdot \ket{v} = R_{V}(g) \ket{v}$ for any $\ket{v} \in V$\footnote{We often alternate between these two equivalent perspectives (action and representations) and their corresponding notations.}. We will always be concerned with \emph{unitary representations} whereby $R_V(g) \in \mathbb{U}(V)$, with $\mathbb{U}(V)$ the group of unitaries on $V$. Then, we have that $R_{V}(g^{-1})=R_V(g)^{-1} = R_V(g)^{\dagger}$. When the group acts on the $n$-qubit space $\hilb$, we write their qubit representation as $R_{\rm q}(g)$.

A representation $R_V$ on a vector space $V$  induces a representation on the operator space ${\rm End}(V)$ defined through $g\cdot O = R_V(g) O R_V(g)^{\dagger}$.
More generally, representations $R_V$ and $R_W$ on spaces $V$ and $W$ induces a representations on the space of maps $\homvw$ through $g \cdot \Gamma = R_W(g) \Gamma R_V(g)^\dagger$. 

\subsubsection{Symmetric and equivariant maps and observables}\label{sec:symm_inv_operators}
Equivariance in Definition~\ref{def:equiv} can take many forms depending on the type of functions considered, and the choice of what constitutes their inputs and outputs. Let us first focus on linear and affine functions.
\begin{definition}[Equivariant linear maps]\label{def:equiv_maps}
Given a group $G$ with representations $R_V$ on $V$ and $R_W$ on $W$, we say that $\Gamma \in \homvw$ is equivariant whenever 
\begin{equation}\label{eq:equiv_maps}
    \forall g: \; R_{W}(g) \Gamma R_{V}(g)^\dagger =  \Gamma.
\end{equation}
The space of such maps is denoted $\homm_{G}(V, W)$.
\end{definition}

Specializing this definition, let us define symmetric and equivariant observables. 
\begin{definition}[Symmetric and equivariant observables]\label{def:symm_operators}
Given a group of symmetry $G$ with representation $R_q$ on $\mathcal{H}$, we say that
an observable $O \in \mathcal{B}$ is symmetric whenever 
\begin{equation}\label{eq:symm_operators}
    \forall g: \; R_q(g) O R_q(g)^\dagger = O \;(\Leftrightarrow [O, R_q(g)]=0).
\end{equation}
An observable $O(x) \in \mathcal{B}$, that depends on some input $x\in \mc{X}$ (on which $G$ also has an action), is equivariant, whenever 
\begin{equation}\label{eq:equiv_operators}
    \forall g, x: \; O(g \cdot x) = R_q(g) O(x) R_q(g)^\dagger.
\end{equation}
\end{definition}
While both display forms of symmetries with respect to $G$, equivariant and symmetric observables differ in that the former depends on some input $x$ while the latter does not.
Nonetheless, through the isomorphism $\hilbop\cong \homm(\mathbb{R}, \hilbop)$, one sees that both symmetric and equivariant observables can be seen as equivariant maps $\homm_{G}(V, \hilbop)$ that have output space $\hilbop$ but different input spaces, with the $1$-dimension $\mathbb{R}$ for the symmetric ones.

To generalize the previous to equivariant \emph{affine} transformations,
recall that an affine transformation $\Gamma \in \affvw$ is composed of a linear term $\Gamma_{\rm lin} \in \homvw$ and a constant term $\Gamma_{0} \in W \equiv  \hom(\mathbb{R, W})$. Given that these two terms are independent, equivariance of $\Gamma$ as per Definition~\ref{def:equiv} translates into a requirement of equivariance for the two terms individually.
This leads us to the following definition.
\begin{definition}[Equivariant affine maps and observables]\label{def:equiv_affine_map}
Given a group $G$ with representations $R_V$ on $V$ and $R_W$ on $W$, we say that $\Gamma \in \affvw$ is equivariant iff there exits $\Gamma_{\rm lin} \in \homm_{G}(V, W)$ and $\Gamma_{0} \in \homm_{G}(\mathbb{R}, W)$ such that for all $\ket{v}\in V$
\begin{equation}\label{eq:equiv_affine_map}
    \Gamma(\ket{v}) = \Gamma_{\rm lin} \ket{v} + \Gamma_0.
\end{equation}
The space of such maps is denoted $\aff_{G}(V,W)$.
When the output space is $\hilbop$,  we say that $\Gamma \in \aff_{G}(V, \hilbop)$ is an equivariant affine observable map.
\end{definition}

In effect, equivariant affine maps generalize symmetric (their input-independent part) and equivariant (their input-dependent part) observables and will be central in the construction of the invariant and equivariant models of Sec.~\ref{sec:symmetric_models}. 

\subsection{Graph problems and the symmetric group}
\begin{figure}
\includegraphics[width=0.48\textwidth]{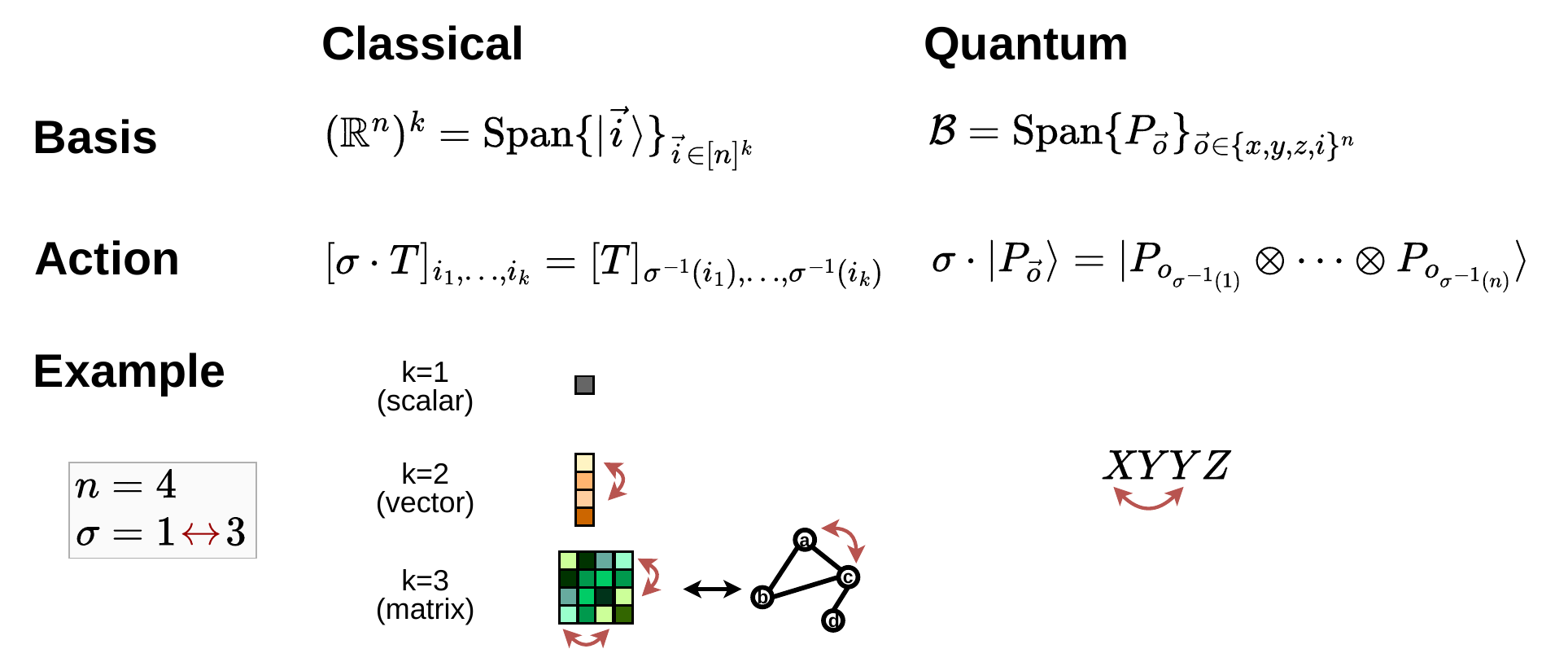}
\vspace{-10pt}\caption{Summary of the actions of $S_n$ on the spaces of interests both classical (order-$k$ tensors) and quantum (operators) spaces. For each we provide the basis and action of a permutation, both formally and through visual examples.
}
\label{fig:actions}
\end{figure}

Our focus is on graph problems: In such problems, labels should not change (invariance) or change in a deterministic fashion (equivariance) upon relabeling of the graph nodes. That is, these problems are equivariant or invariant under the action of the symmetric group. 
 In this section we review background of the symmetric group (Sec.~\ref{sec:graph_and_sn}), then present its action on graphs and other spaces of interests. 
In particular, given that QML models interface classical and quantum spaces, we need to define action of permutations on both classical (Sec.~\ref{sec:act_classical}) and quantum spaces (Sec.~\ref{sec:act_quantum})\footnote{Mathematically, these are only different representation spaces for the symmetric group, still we make this, somewhat arbitrary, distinction due to the way the corresponding operations on these spaces are implemented.}.
 A summary of these actions is provided in Fig.~\ref{fig:actions}.
 In all that follows, we restrict our study to quantum models acting on a number of qubits equal to the number of nodes in the underlying graph: \emph{$n$-node graphs are encoded as $n$-qubit states}.

\subsubsection{Graphs and the symmetric group}\label{sec:graph_and_sn}
A graph $x$ consists of a set $\mc{V}(x)$ of $n$ vertices and a set of (potentially weighted and directed) edges $\mc{E}(x)$. These are equivalently identified through their adjacency matrices such that we can take inputs to be matrices, $x \in \mathbb{R}^{n \times n}$, with entries $x_{i,j}$ the weight of the edge from node $i$ to $j$. When $F$ additional nodes or edges \emph{features} are available, as is often the case in ML problems, $x \in \mathbb{R}^{n \times n \times F}$ with the additional dimensions accounting for the features. 
In the following, if not otherwise stated, when talking about graphs we mean \emph{featureless} graphs. 

Given an $n$-node graph, the underlying group of symmetry is the symmetric group $S_n$ consisting of $n!$ permutations over the set of indices $[n]:=\set{1, \hdots, n}$. For a permutation $\sigma \in S_n$, denote as $\sigma(i) \in [n]$ its action on an index, and given a vector of index $\vec{i}=(i_1, \hdots, i_m)\in [n]^m$ define $\sigma(\vec{i}):=(\sigma(i_1), \hdots, \sigma(i_m))$.
Recall that $S_n$ is $m$-transitive for any $m\leq n$: Given any two vectors of $m$ \emph{pairwise distinct} indices, $\vec{i}=(i_1, \hdots, i_m)$ and $\vec{i}' = (i'_1, \hdots, i'_m)$, there always exists a permutation $\sigma$ such that $\sigma(\vec{i})=\vec{i}'$.

Further recall that any $\sigma \in S_n$ can be decomposed as a transposition $(1 \leftrightarrow j_1)$ with $j_1 \in [n]$ and a permutation $\sigma_2 \in S_{n-1}$ that only permutes indices $j_2 \in [n]_2 :=\{2, \hdots, n\}$. Similarly $\sigma_2$ can be decomposed as a transposition $(2 \leftrightarrow i_2)$ with $i_2 \in [n]_2$ and a permutation $\sigma_3 \in S_{n-2}$ acting non trivially only on $j_3 \in [n]_3 :=\{3, \hdots, n\}$. Through recursion one obtains a decomposition of any permutation $\sigma$ as a product of $n-1$ transpositions of the form
\begin{equation*}
    \sigma = \prod_{i=1}^{n-1} (i \leftrightarrow j_i), \, \textrm{with  } j_i \in [n]_i.
\end{equation*}
This decomposition essentially expresses $\sigma$ using the \textit{set isomorphism} $S_n \cong S_n/S_{n-1} \times \cdots S_2$, with the $(i \leftrightarrow j_i)$ a choice of transversals for $S_{n-i}/S_{n-i-1}$.

As is often the case when dealing with the symmetric group, combinatorics notions are helpful. Recall that a set of $k$ indices $\set{i_1, \dots, i_k}$ can be partitioned in $b(k)$ different ways, with $b(k)$ being the Bell numbers. An example of such partition, into three non-empty subsets, is $\{\{i_1\}, \{i_2\}, \{i_3, \dots, i_k\} \}$ . The number of partitions of $k$ indices into $k'$ non empty subset is given by the Stirling number of the second kind ${k \brace k'}$, with $\sum^k_{k'=1} {k \brace k'} = b(k)$.

\subsubsection{Action on classical vector spaces}\label{sec:act_classical}
Permutations acts on a graph by relabeling its vertices, or equivalently, on its adjacency matrice $x$ by permuting \emph{simultaneously} its rows and columns. More generally, consider an order-k tensors $T \in \mathbb{R}^{n^k}$, with entries indexed through a vector of indices $\vec{i}\in [n]^k$ such as $[T]_{\vec{i}}$ or $[T]_{i_1,\hdots, i_k}$ (we often drop the brackets).
The action of any $\sigma \in S_n$ is defined through
\begin{equation}\label{eq:action_Sn_Tensors}
[\sigma\cdot T]_{i_1,\hdots,i_k} = [T]_{\sigma^{-1}(i_1), \hdots, \sigma^{-1}(i_k)}.\footnote{
For graphs (or more generally tensors) with features, permutations only act on the non-feature indices. For instance, for $k=2$ and $f$ the feature index, $[\sigma\cdot T]_{i_1,i_2, f} = [T]_{\sigma^{-1}(i_1), \sigma^{-1}(i_2), f}$.}
\end{equation}
In addition to $k=2$ (matrices or graphs), for $k=0$ (scalar) the action is trivial: $\sigma \cdot T = T  \in \mathbb{R}$, while for $k=1$ (vectors) it simply permutes elements of the vector. 

We use \emph{elementary tensors} as an orthonormal basis of the space of order-$k$ tensors:
\begin{equation}\label{eq:basis_orderk}
    \mathbb{R}^{n^k} = {\rm Span} \set{\dketsp{\vec{i}}= \dket{i_1, \hdots, i_k}}_{\vec{i}\in[n]^k}
\end{equation}
with $\dketsp{\vec{i}}$ being an order-$k$ elementary tensor that has value one at coordinate $\vec{i}$ and zero otherwise. As an example, for $k=2$, $\dket{i_1,i_2}$ is the $n \times n$ matrix with a $1$ value at row $i_1$ and column $i_2$, and $0$ elsewhere. From Eq.~\eqref{eq:action_Sn_Tensors}, the action of a permutation on these tensors reads
\begin{equation}\label{eq:action_elementary}
    \sigma\cdot \dketsp{\vec{i}} = \dket{\sigma(\vec{i})} = \dket{\sigma(i_{1}), \hdots, \sigma(i_{k})},
\end{equation}
showing that any $\sigma$ will always map one elementary tensor to another.
 For $k=0$, the $1$-dimensional vector space has basis denoted as $\dket{\mathbf{0}}$ (the unit scalar), any permutation $\sigma$ acts trivially on it: $\sigma \cdot \dket{\mathbf{0}} = \dket{\mathbf{0}}$.

\subsubsection{Action on quantum vector spaces}\label{sec:act_quantum}
Moving to quantum spaces, the action of a permutation $\sigma \in S_n$ on any n-qubit separable state is defined through
\begin{align}\label{eq:action_Sn_qubits}
\begin{split}
    \sigma \cdot \big( \ket{\psi_1} \otimes \hdots \otimes \ket{\psi_n} \big) &:= R_{\rm q}(\sigma) \big( \ket{\psi_1} \otimes \hdots \otimes \ket{\psi_n} \big) \\
    &= \ket{\psi_{\sigma^{-1}(1)}} \otimes \hdots \otimes \ket{\psi_{\sigma^{-1}(n)}},
\end{split}
\end{align}
where $\ket{\psi_j}$ denotes the state of the $j$-th qubit.
Recall that $R_{\rm q}(\sigma)$ is the unitary representation of $\sigma$ onto the $n$-qubit state.
Eq.~\eqref{eq:action_Sn_qubits} is extended by linearity to any state $\ket{\psi} \in \hilb$ and to any $O\in \hilbop$ (Sec.~\ref{sec:notations}) through
\begin{equation}\label{eq:action_Sn_operators}
    \sigma \cdot O := R_{\rm q}(\sigma) O R_{\rm q}(\sigma)^{\dagger}.
\end{equation}
As a basis of the observable space we resort to Pauli strings,
\begin{equation}\label{eq:basis_op}
    \hilbop = {\rm Span} \set{\dket{P_{\vec{o}}} := \dket{P_{o_1} \otimes \hdots \otimes P_{o_n}}}_{\vec{o}\in\{x, y, z, i\}^n},
\end{equation}
where each Pauli string $P_{\vec{o}}$ is a tensor product of $n$ Pauli operators $P_{o} \in \set{X, Y, Z, I}$ acting on the $o$-th qubit.
The action of a permutation on a Pauli string amounts to permuting the  qubits on which Pauli operators act:
\begin{equation}\label{eq:action_Sn_Paulis}
    \sigma \cdot \dket{P_{o_1} \otimes \hdots \otimes P_{o_n}} = \dket{P_{o_{\sigma^{-1}(1)}} \otimes \hdots \otimes P_{o_{\sigma^{-1}(n)}}}.
\end{equation}
This is extended by linearity to any $O \in \hilbop$.

Going further, \emph{order-$k$ tensor of observables} are defined as a collection of $n^k$ observables arranged as a tensor $\mathbf{A}$ with entries $[\mathbf{A}]_{\vec{i}}\in \mc{B}$ indexed by the vector $\vec{i} \in [n]^k$.
As per the action of $S_n$ on real valued tensors~\eqref{eq:action_Sn_Tensors}, action on tensors of observables  is given by
\begin{equation}\label{eq:act_operators_tensor}
    [\sigma\cdot \mathbf{A}]_{i_1,\hdots,i_k} = [\mathbf{A}]_{\sigma^{-1}(i_1), \hdots, \sigma^{-1}(i_k)}.
\end{equation}
Given a state $\rho$, we denote by $\mathbf{A}[\rho] \in \mathbb{R}^{n^k}$ the order-$k$ tensors of expectation values $\Tr[ \mathbf{A}_{\vec{i}} \rho] \in \mathbb{R}$. In turn, the action of $S_n$ on $\mathbf{A}$~\eqref{eq:act_operators_tensor} induces an action on  $\mathbf{A}[\rho]$ consistent with the action on real tensors~\eqref{eq:action_Sn_Tensors}.

\section{A toolbox for GQML on graphs}\label{sec:toolbox}
\begin{figure}
\includegraphics[width=0.5\textwidth]{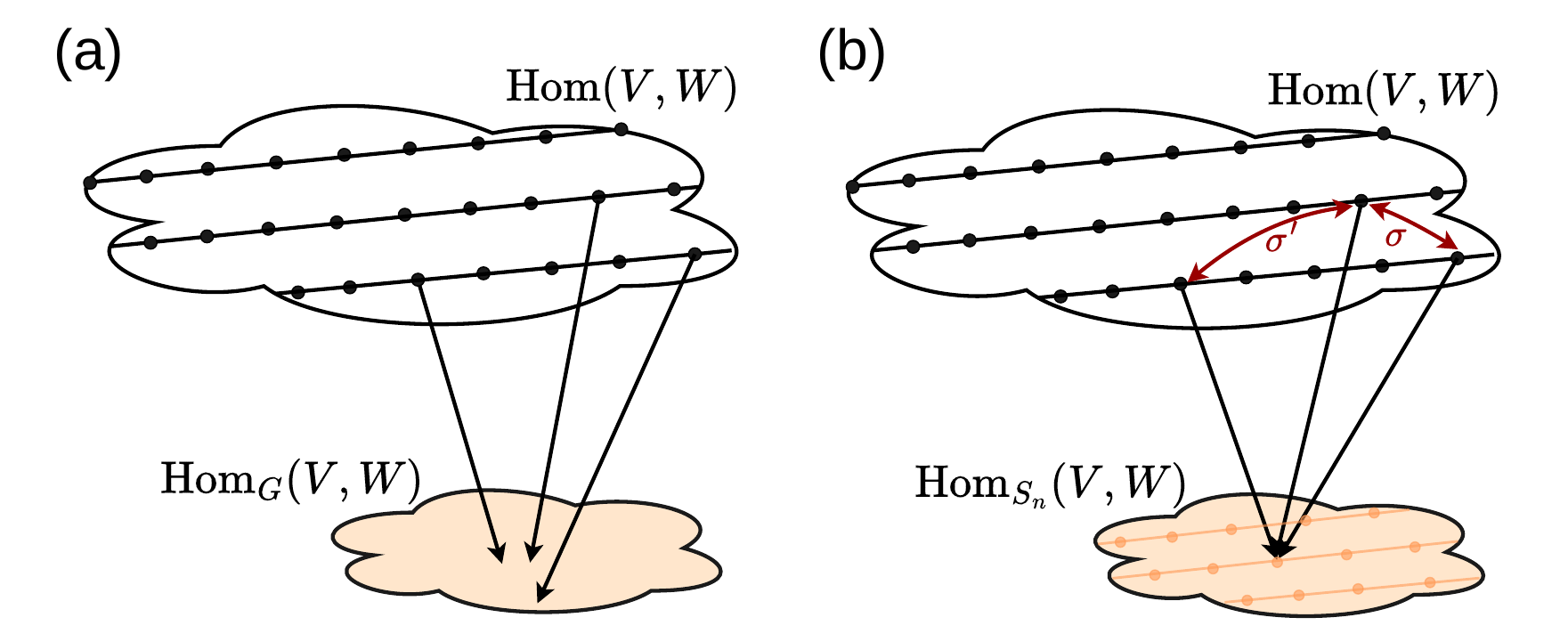}
\vspace{-10pt}\caption{\textbf{(a)} Symmetrizing~\eqref{eq:def_twirl} all elements of a basis of linear maps $\textrm{Hom}(V,W)$, one obtain a set of maps spanning the space of all equivariant maps $\textrm{Hom}_G(V,W)$. \textbf{(b)} While in general this procedure is not scalable, this becomes tractable whenever the basis is closed under the action of the group symmetry (ie any element of the basis is mapped to another one under group action). This is the case for our choice of bases for $S_n$, and we use this strategy to identify all the equivariant linear maps necessary to our models.
}
\label{fig:twirl}
\end{figure}

\begin{table*}
\begin{tabular}{lllll}
\hline\hline
V & W & Interpretation& Reference & Defining properties ($\forall \sigma \in S_n$) \\ [0.5ex]
\hline 
$\mathbb{R}$      & $\hilbop$  & Symmetric observables& Eq.~\eqref{eq:symm_operators} & $A=R_{\sigma}AR_{\sigma}^{\dagger}$  \\ 
$\mathbb{R}^{n \times n}$  & $\hilbop$ & Equivariant observables & Eq.~\eqref{eq:equiv_operators}& $A(\sigma \cdot x) = R_{\sigma}A(x)R_{\sigma}^{\dagger}$ \\
$\hilbop$  & $\mathbb{R}$& Invariant measurements (single observable)& Eq.~\eqref{eq:inv_meas}& ${\rm Tr}[R_{\sigma}\rho R_{\sigma}^{\dagger} A] = {\rm Tr}[\rho A]$\\
$\hilbop$  & $\mathbb{R}^n$ & Equivariant measurements (vector of observables)& Eq.~\eqref{eq:equiv_vec_meas} & ${\rm Tr}[R_{\sigma}\rho R_{\sigma}^{\dagger} \mathbf{A}_i] = {\rm Tr}[\rho \mathbf{A}_{\sigma^{-1}(i)}]$\\
$\hilbop$  & $\mathbb{R}^{n \times n}$ & Equivariant measurements (matrix of observables)& Eq.~\eqref{eq:equiv_mat_meas} & ${\rm Tr}[R_{\sigma}\rho R_{\sigma}^{\dagger} \mathbf{A}_{i,j}] = {\rm Tr}[\rho \mathbf{A}_{\sigma^{-1}(i), \sigma^{-1}(j)}]$\\[0.5ex]
\hline
\end{tabular}
\caption{Example of families of equivariant linear maps $A \in {\rm Hom}_{S_n}(V, W)$ for different choices of input $V$ and output domain $W$ together with their interpretations and properties. Here, $\sigma \in S_n$ denotes an arbitrary permutation with $R_{\sigma}:=R_q(\sigma)$ its representation on $\hilb$, and $x\in \mathbb{R}^{n \times n}$ denotes an arbitrary input graph. As discussed further in the main text, in the context of measurements (last three rows) the maps can be associated to either a single observable $A$ or a tensor of observables $\mathbf{A}$.} 
\label{tab:equiv_maps_interpr}
\end{table*}
In this section, we characterize the primitives composing GQML models, namely equivariant linear and non-linear maps.
The key to identifying linear ones is the concept of symmetrization that is reviewed in Sec.~\ref{sec:twirl} and depicted in Fig.~\ref{fig:twirl}. Although in general intractable, in our situation, this yields a systematic and efficient procedure, the results of which are reported in Sec.~\ref{sec:equiv_linear_maps}.
Finally in Sec.~\ref{sec:equiv_nonlinear_maps} we characterize the equivariant non-linear maps at our disposal.

\subsection{Symmetrization}\label{sec:twirl}
Equivariant linear maps, as per Definition~\ref{def:equiv_maps}, can be identified through \emph{symmetrization}: given any linear map $\Gamma \in \homvw$ from $V$ to $W$, where $\sigma \in S_n$ admits a representation $R_V(\sigma)$ and $R_W(\sigma)$, its symmetrization (or twirl) is defined as the average under the action of $S_n$:  
\begin{equation}\label{eq:def_twirl}
\mc{T}(\Gamma) = \frac{1}{n!}\sum_{\sigma \in S_n} \sigma \cdot \Gamma = \frac{1}{n!}\sum_{\sigma \in S_n} R_W(\sigma) \Gamma R_V(\sigma)^\dag.
\end{equation}
This has been used  in the context of QML~\cite{meyer2023exploiting} for symmetric operators (when $V=W$).

The maps resulting from Eq.~\eqref{eq:def_twirl} are equivariant: $\mc{T}(\Gamma) \in \homm_{S_n}(V, W)$ for any $\Gamma$.
In addition, one can show that
the symmetrization is an \emph{orthogonal projection} onto the subspace $\homm_{S_n}(V, W)$~\cite{fulton2013representation}.
Then, for bases $B_V=\{\ket{v}\}$ of $V$ and $B_W=\{\ket{w}\}$ of $W$, we get
\begin{equation}\label{eq:twirl_all}
    \homm_{S_n}(V, W) = {\rm Span} \set{\mc{T}\left( \ket{w}\bra{v} \right)}_{\ket{v} \in B_V, \ket{w} \in B_W}
\end{equation}
That is, by symmetrizing all elements of the basis $B_{V, W}=\set{\ket{w}\bra{v}}$ of $\homvw$, we obtain a set of maps that are guaranteed to span the space of equivariant maps. In the following, we detail how this yields a scalable strategy for the identification of linear equivariant maps (Sec.~\ref{sec:systematic_strat}) together with the interpretation and realization of such maps (Sec.~\ref{sec:interpretation}). 

\subsubsection{Systematic strategy}\label{sec:systematic_strat}
Eq.~\eqref{eq:twirl_all} yields a systematic way to characterize $\homm_{S_n}(V, W)$. 
However, in general, this would incur intractable computations, as individual twirls require summing over $|S_n|=n!$ terms and need to be performed for all $\dim(V) \times 
\dim(W)$ basis elements. Furthermore, the resulting symmetrized maps need not be orthogonal (as $\mc{T}$ is a projector), yielding an overcomplete basis, complicating the task of parameterizing equivariant spaces\footnote{To parameterize an observable $O_{\theta} \in \homm_{S_n}(V, W)$ one would assign a real parameter $\theta_i$ to each element of its basis.}.

Nonetheless, for our purpose  -- for input and output spaces $V,W \in \{\mathbb{R}^{n^k}, \mc{B}\}$ and corresponding orthogonal bases defined in Eqs.~\eqref{eq:basis_orderk} and~\eqref{eq:basis_op} -- such a strategy can be shown to be efficient. 
The key to this is the fact that the action of $\sigma$ onto any basis elements of $V$ (or $W$) produces another basis elements of $V$ (or $W$) rather than linear combination of such,
as per Eqs.~\eqref{eq:action_elementary} and~\eqref{eq:action_Sn_Paulis}. Then, $\sigma \cdot \ket{w}\bra{v} = \ket{w'}\bra{v'} \in B_{V,W}$ and we can \emph{partition} $B_{V,W}$ in terms of equivalence classes defined as
\begin{equation}\label{def:equiv_classes}
    \mc{C}_{\ket{w}\bra{v}} = \set{\sigma \cdot \ket{w}\bra{v}}_{\sigma \in S_n}.
\end{equation}
In turn, from Eq.~\eqref{eq:def_twirl} we see that symmetrization of a given $\ket{w}\bra{v}$ is, up to a scaling constant, the average of all elements belonging to its equivalence class $\mc{C}_{\ket{w}\bra{v}}$. 
Hence, we get a basis of the equivariant maps as
\begin{equation}
    \homm_{S_n}(V, W) = {\rm Span} \bigg\{ \sum_{\ket{w}\bra{v} \in \mc{C}} \ket{w}\bra{v} \bigg\}_{\mc{C}},
\end{equation}
which is \emph{orthogonal} given the orthogonality of $B_{V,W}$. 
Overall, this shows that identifying an orthogonal basis of $\homm_{S_n}(V, W)$ only requires identifying all the equivalence classes~\eqref{def:equiv_classes} (or equivalently, a representative for each class). 

\subsubsection{Interpretation}\label{sec:interpretation}
The previous approach yields efficient computations such that we can identify all equivariant linear maps of interest. 
Before proceeding to concrete instantiations of this strategy, we provide the interpretation and definite properties of the equivariant maps $\Gamma \in \homm_{S_n}(V, W)$ satisfying Definition.~\ref{def:equiv_maps} for various choices of input spaces $V$ and output spaces $W$. These are summarized in Table~\ref{tab:equiv_maps_interpr}. 
Note that $\homm_{S_n}(V, W) \equiv \homm_{S_n}(W, V)$ such that some of the families of equivariant maps discussed are effectively the same (ie isomorphic to another) and only differ in their use and interpretation.

\para{Case I $(W = \hilbop)$} 
The maps $\Gamma$ yield observables.
When $V = \mathbb{R}$ (first row), we obtain the symmetric observables~\eqref{eq:symm_operators}, while
when $V=\mathbb{R}^{n \times n}$ (second row), we obtain the equivariant ones~\eqref{eq:symm_operators}. 
In both cases, the output of the maps can be used as generators of unitary gates.

\para{Case II $(V = \hilbop)$}
Considering now maps from observables (such as density matrices) to real-valued tensors, these can be realized through measurements. 
For instance, measuring an observable $A$ implements a map $\hilbop \mapsto \mbb{R}$ through $\rho \mapsto {\rm Tr}[A \rho]$. 
In fact, \emph{any} linear map $\Gamma: \hilbop \mapsto \mbb{R}$ can uniquely be associated with an observable $A_{\Gamma}\in \hilb$.
Thus, for $W=\mbb{R}$ (third row), each equivariant map $\Gamma$ corresponds to an observable $A_{\Gamma}$ that satisfies 
\begin{equation}\label{eq:inv_meas}
    {\rm Tr}[A_{\Gamma} R_q(\sigma)\rho R_q(\sigma)^\dagger]={\rm Tr}[A_{\Gamma} \rho ],
\end{equation}
that is, to an \emph{invariant measurement}.

For $W=\mathbb{R}^{n^k}$ with $k \geq 1$ (fourth and fifth row), one rather identifies each element of the output (tensor) of a map to an observable. That is, \emph{any} map $\Gamma$ can uniquely be associated with an order-$k$ tensor of observables $\mathbf{A}_{\Gamma}$ as defined in Sec.~\ref{sec:act_quantum}.
In turn, equivariance of the map~\eqref{eq:equiv_maps} entails equivariance of the tensor of observables defined as follows.
\begin{definition}[Equivariant tensor of observables]\label{def:equiv_operators_tensor}
We say that an order-$k$ tensor of observables $\mathbf{A}$ is $S_n$-equivariant, whenever for all $\sigma \in S_n$ these satisfy
\begin{equation}\label{eq:equiv_operators_tensor}
    R_q(\sigma)^\dagger \left(\mathbf{A}_{i_1, \hdots, i_k}\right) R_q(\sigma) = \mathbf{A}_{\sigma^{-1}(i_1), \hdots, \sigma^{-1}(i_k)}.
\end{equation}
\end{definition}

Eq.~\eqref{eq:equiv_operators_tensor} is understood as a requirement that the action of $S_n$ on each individual observables (LHS) matches the action on the tensor (RHS) as per Eq.~\eqref{eq:equiv_operators_tensor}. We stress that this equality does not hold for arbitrary tensor of observables.

For $k=0$, we recover Eq.~\eqref{eq:inv_meas} from Eq.~\eqref{eq:equiv_operators_tensor}. For $k=1$ (fourth row), any map $\Gamma \in \homm_{S_n}(\mc{B}, \mathbb{R}^{n})$ can be associated to a vector of observables $\mathbf{A}_{\Gamma}$ satisfying:
\begin{equation}\label{eq:equiv_vec_meas}
    {\rm Tr}\left[[\mathbf{A}_{\Gamma}]_i R_q(\sigma)\rho R_q(\sigma)^\dagger\right]={\rm Tr}\left[[\mathbf{A}_{\Gamma}]_{\sigma^{-1}(i)} \rho \right].
\end{equation}
 For $k=2$ (fifth row), $\Gamma \in \homm_{S_n}(\mc{B}_n, \mathbb{R}^{n \times n})$ are associated to matrices of observables $\mathbf{A}_{\Gamma}$ satisfying:
\begin{equation}\label{eq:equiv_mat_meas}
    {\rm Tr}\left[[\mathbf{A}_{\Gamma}]_{i,j} R_q(\sigma)\rho R_q(\sigma)^\dagger\right]={\rm Tr}\left[[\mathbf{A}_{\Gamma}]_{\sigma^{-1}(i),\sigma^{-1}(j)} \rho \right].
\end{equation}

To summarize, maps $\Gamma \in \homm_{S_n}(\mc{B}_n, \mathbb{R}^{n^k})$ are in one-to-one correspondence with equivariant tensor of observables (Definition~\ref{def:equiv_operators_tensor}) which when used for measurements yield invariance and equivariance constraints on the tensors of expectation values measured as per Eqs.~\eqref{eq:inv_meas},~\eqref{eq:equiv_vec_meas} and~\eqref{eq:equiv_mat_meas}.
As we will discuss further, the latters are key to lift invariant QML models to equivariant ones. 
Finally, we note that the concepts of (equivariant) tensor of operators appears in the context of the Quantum Schur Transform~\cite{bacon2005quantum} and is similar to the \emph{representation operators} that have been used to generalize equivariance in the context of convolutional NNs~\cite{cohen2017steerable}. One notable difference is that in the previous, operators are typically arranged as a vector that is an irreducible representation for $S_n$. Here, we do not insist on the irreducible nature of our tensors. Rather, such tensors are natural in that they allow to interface quantum models to classical ones as discussed in Sec.~\ref{sec:symmetric_models}.

\subsection{Linear equivariant maps}\label{sec:equiv_linear_maps}
Applying the symmetrization strategy of  Sec.~\ref{sec:twirl}, we provide a comprehensive characterization of all equivariant linear maps $\homm_{S_n}(V, W)$ relevant to our models. Main results are reported in Table~\ref{tab:equiv_maps} and are discussed in the following.
We start by presenting the maps involving only classical spaces (Sec.~\ref{sec:ex_classical} and first row of Table~\ref{tab:equiv_maps}) and then proceed to the maps mixing classical and quantum spaces (Sec.~\ref{sec:ex_cq} and second row of Table~\ref{tab:equiv_maps}). For the latter we further show how to restrict the observable space to account for implementation desiderata (Sec.~\ref{sec:special} and third to seventh row of Table~\ref{tab:equiv_maps}).
In all cases, we start by elaborating on simple working examples from which we build towards general results.

\subsubsection{Classical spaces}\label{sec:ex_classical}
\para{Case I $(V=\mathbb{R}^{n \times n}, W=\mathbb{R})$}
The basis of the linear maps is $B_{V,W} = \{\dket{\mathbf{0}}\dbra{i_1, i_2}\}$ with $(i_1,i_2) \in [n]^2$. 
Given Eq.~\eqref{eq:action_elementary} and the $1$-transitivity of $S_n$, there always exists $\sigma$ that maps any $\dket{\mathbf{0}}\dbra{i_1,i_1}$ to any other  $\dket{\mathbf{0}}\dbra{i_1',i_1'}$. Resorting the $2$-transitivity of $S_n$, there always exists $\sigma$ that maps any $\dket{\mathbf{0}}\dbra{i_1,i_2}$ to any other  $\dket{\mathbf{0}}\dbra{i_1',i_2'}$ provided that $i_1 \neq i_2$ and $i'_1 \neq i'_2$. However, no permutation can map $\dket{\mathbf{0}}\dbra{i_1,i_1}$ to $\dket{\mathbf{0}}\dbra{i'_1,i_2' \neq i'_1}$. The basis of maps thus partitions in $2$ equivalence classes denoted as
\begin{equation*}
\mc{C}_0 \equiv \{\{i_1, i_2\}\}, \, \textrm{and} \,\mc{C}_1 \equiv \{\{i_1\}, \{i_2\}\},
\end{equation*}
 to signify that the two indices take the same (distinct) values for the first (second) equivalence classes.
Correspondingly, the equivariant maps are spanned by the two orthogonal equivariant maps
\begin{equation}\label{eq:expl_map_20_0}
    \Gamma_0 = \sum_{i_1 \in [n]} \dket{\mathbf{0}}\dbra{i_1,i_1}, \;\; \Gamma_1 = \sum_{i_1 \neq i_2 \in [n]^2} \dket{\mathbf{0}}\dbra{i_1,i_2}, 
\end{equation}
that perform a sum over all  diagonal and all off-diagonal elements, respectively. 

\para{Case II $(V=\mathbb{R}^{n \times n}, W=\mathbb{R}^{n \times n})$} 
Starting with a basis $B_{V,W}=\set{\dket{i_1, i_2}\dbra{i_3, i_4}}$, we get $15$ equivalence classes, that are all the different ways of tying values of the indices or, equivalently, all the partitions of $4$ symbols. As an example for the equivalence class $\mc{C}=\{\{i_1, i_4\}, \{i_2, i_3\} \}$ we get the equivariant map 
\begin{equation}\label{eq:expl_map_22_0}
    \Gamma = \sum_{i_1 \neq i_2 \in [n]} \dket{i_1, i_2}\dbra{i_2,i_1},
\end{equation}
that transposes the off-diagonal part of the input matrix and remove the diagonal part. 

\para{General case $(V=\mathbb{R}^{n^k}, W=\mathbb{R}^{n^{k'}})$}
The basis $B_{V,W}=\set{\dket{i_1, \hdots, i_{k'}}\dbra{i_{k+1}, \hdots, i_{k+k'}}}$ now contains $k+k'$ indices. The equivalence classes are all partitions of these symbols and can be translated into explicit maps as before for Eqs.~\eqref{eq:expl_map_20_0} and~\eqref{eq:expl_map_22_0}.
The number of partitions on $k+k'$ indices is the \emph{Bell number} denoted $b(k+k')$\footnote{Here and elsewhere, we assume that $n \geq k+k'$ (or $n\geq k$). When not the case, we cannot map $l>n$ partitions to $n$ distinct values, and the Bell numbers would be replaced by the $n$-th restricted ones: $b(k,n):=\sum^n_{k'=1} {n \brace k'}$~\cite{pearce2022connecting}.} which is reported in Table~\ref{tab:equiv_maps} (G1, second line). 
Note that these maps were identified in Ref.~\cite{maron2018invariant,pearce2022connecting} and their advantage in the context of GNNs is shown in Ref.~\cite{maron2018invariant}. The appeal of the explicit symmetrization approach developed here lies in that it can equivalently be used it for maps from (to) classical tensor spaces to (from) quantum spaces as demonstrated in the next section.

\begin{table*}[t!]
\centering
\begin{tabular}{lll lll}
\hline\hline
V & W & Dimension & Use cases & Label \\ [0.5ex]
\hline
$\mathbb{R}^{n^{k}}$      & $\mathbb{R}^{n^{k'}}$  & $b(k+k')$        & General equivariant layers of GNNs~\cite{maron2018invariant} & G1\\ 
$\mc{B}$  & $\mathbb{R}^{n^k}$ & $\sum^k_{k'=1} {k \brace k'} 4^{k'} {\rm Te}(n+1-k')$   & General encoding and measurements for GQML on graphs & G2\\ [1ex]
\hline
$\mathbb{R}$  & $\mc{B}_{\{I, Z\}}$ & $n+1$ & Input-independent observables in the Z-basis & M1\\
$\mathbb{R}^{n \times n}$  & $\mc{B}_{\{I, Z\}}$ & $6n-4$   & Input-dependent observables in the Z-basis & M2 \\
$\mc{B}_{\{I, Z\}}$  & $\mathbb{R}^{n}$   & $2n$ & Input-independent observables in the Z-basis & M3\\
$\mathbb{R}$  & $\mc{B}_{l\leq2}$ & $10$   & Input-independent generator with bodyness $l\leq 2$& E1 \\
$\mathbb{R}^{n \times n}$  & $\mc{B}_{l\leq2}$   & $49$ & Input-dependent generator with bodyness $l\leq 2$ & E2\\[0.5ex]
\hline
\end{tabular}
\caption{Equivariant linear maps for relevant input ($V$) and output ($W$) domains. We report the dimension of the equivariant spaces $\homm_{S_n}(V, W)$ with their use cases. Bell, Tetrahedral and Stirling numbers of the second kind are denoted as $b(\cdot)$, $\textrm{Te}(\cdot)$ and ${\cdot \brace \cdot}$, respectively. General results are provided (G1-G2) together with more specialized ones (M1-M3, E1-E2) for subspaces of observables $\mc{B}_{\{I, Z\}}$ or $\mc{B}_{l\leq2}$, spanned by Pauli strings $\{I,Z\}^{\otimes n}$ or Pauli strings that are no more than $2$-local, respectively.}
\label{tab:equiv_maps}
\end{table*}
\subsubsection{Quantum spaces }\label{sec:ex_cq}
\para{Case I $(V=\hilbop, W=\mathbb{R})$}
The basis of the linear maps is $B_{V,W} = \{\dket{\mathbf{0}}\dbra{P_{\vec{o}}}\}$ with $\vec{o} \in \{x,y,z,i\}^n$.
Let us denote $k(\vec{o})\equiv [k_x(\vec{o}), k_y(\vec{o}), k_z(\vec{o}), k_i(\vec{o})] \in \mathbb{N}^4$ the composition of the corresponding Pauli string: the number of individual $X$, $Y$, $Z$ and $I$ appearing in the string. 
Any Pauli string $P_{\vec{o}}$ can be mapped to any other $P_{\vec{o}^\prime}$ provided that $k(\vec{o})= k(\vec{o}\,')$. However, Pauli strings with different compositions cannot be mapped to one another through permutations. Hence, equivalence classes $\mc{C}_{\vec{k}}$ are identified by all compositions $\vec{k}\in \mbb{N}^4$ satisfying $\|\vec{k}\|_1:= k_x+k_y+k_z+k_i=n$ (we say that such compositions are valid and denote them through $\vec{k} \vdash n$). To each composition $\vec{k}$ we associate the (symmetric) observable
\begin{equation}\label{eq:symm_sn_op}
    S_{\vec{k}} = \sum_{\vec{o}: k(\vec{o}) = \vec{k}} P_{\vec{o}},
\end{equation}
that is a sum of all Pauli strings with the same composition.
Then, it follows that the space of linear equivariant maps is spanned as:
\begin{equation}\label{eq:meas_inv_op}
        \homm_{S_n}(\hilbop, \mathbb{R}) = {\rm Span} \set{\dket{\mathbf{0}}\dbra{S_{\vec{k}}}}_{\vec{k} \vdash n}
\end{equation}
As discussed in Sec.~\ref{sec:interpretation}, each of these maps is interpreted as a measurement of the observable $\dket{\vec{k}}$. 

\para{Case I bis $(V=\mathbb{R}, W=\hilbop)$} As discussed earlier $\homm_{S_n}(V, W) \equiv \homm_{S_n}(W, V)$, and Eq.~\eqref{eq:meas_inv_op} becomes
\begin{equation}
        \homm_{S_n}(\mathbb{R}, \hilbop) = {\rm Span} \set{\dket{S_{\vec{k}}}\dbra{\mathbf{0}}}_{\vec{k} \vdash n}
\end{equation}
where now each of the maps is interpreted as a symmetric observable~\eqref{def:symm_operators}, rather than a measurement.

The symmetric observables~\eqref{eq:symm_sn_op} appear in Refs.~\cite{toth2010permutationally, schatzki2024theoretical,anschuetz2023efficient,sauvage2024classical}: Their dimension is the Tetrahedral number ${\rm Te}(n+1)\in \mc{O}(n^3)$ obtained by counting the number of valid compositions $\vec{k}\vdash n$. 
Examples of these for $k_i=n-1$ and $k_x=1$ or $k_y=1$ or $k_z=1$ are, respectively,
\begin{align}\label{ex:symm_operators}
    \begin{split}
    & S_X:=\sum_{i=1}^n X_i, \,
    S_Y:=\sum_{i=1}^n Y_i,
    \, \textrm{and}\; S_Z:=\sum_{i=1}^n Z_i.
    \end{split}
\end{align}

\para{Case II $(V=\hilbop, W=\mathbb{R}^n)$}
As before, we wish to partition the basis of all linear maps, $B_{V,W} = \{\dket{1}\dbra{P_{\vec{o}}}\}$ with $\vec{o} \in \{x,y,z,i\}^n$ and $i \in [n]$, into equivalence classes. Noting that any $\dket{i\neq 1}\dbra{P_{\vec{o}}}$ can always  be mapped to $\dket{1}\dbra{(1 \leftrightarrow i) \cdot P_{\vec{o}}}$, we only need to partition $\set{\dket{ 1}\dbra{P_{\vec{o}}}}$. Only permutations $\sigma_2 \in S_{n-1}$, acting on the last $n-1$ qubits, can map between elements of this set.
Given that they preserve the composition of the Pauli strings on the last $n-1$ qubits, all equivalence classes are labeled by a pair $(P, \vec{k})$ with a Pauli operator $P$ acting on one (the first) qubit, and a valid composition $\vec{k}\vdash (n-1)$ over the $n-1$ remaining qubits, with a total of $4 \times {\rm Te}(n)$  equivalence classes.

For the equivariant maps,
take $\Gamma_{\rm r} = \dket{1}(\dbra{P} \otimes  \dbra{P_{\vec{o}}}$), with any $P_{\vec{o}}$ satisfying $\vec{k}(P_{\vec{o}}) = \vec{k} \vdash (n-1)$, as a representative of the class $(P, \vec{k})$. Upon symmetrization~\eqref{eq:def_twirl},
\begin{align}
\begin{split}
    \mathcal{T}(\Gamma_{\rm r}) &= \sum_{\sigma \in S_n} \sigma \cdot \Gamma_{\rm r} = \sum_{i \in [n]} (1 \leftrightarrow i ) \cdot  \sum_{\sigma' \in S_{n-1}} \sigma' \cdot \Gamma_{\rm r} \\
    &= \sum_{i \in [n]} (1 \leftrightarrow i ) \cdot (\dket{1}\dbra{P, S_{\vec{k}}}),
    \end{split}
\end{align}
where we used the decomposition $\sigma = (1 \leftrightarrow i) \cdot \sigma'$ with $\sigma' \in S_{n-1}$ (Sec.~\ref{sec:graph_and_sn}), and denoted as $\dket{P, S_{\vec{k}}}= \dket{P} \otimes \dket{S_{\vec{k}}}$ with $P$ acting on the first qubit and $S_{\vec{k}}$~\eqref{eq:symm_sn_op} on the remaining ones. 
As per Sec.~\ref{sec:interpretation}, this map is identified as a vector of $n$ observables $\mathbf{A} = (A_1, \hdots, A_n)$ with $A_1 = P \otimes S_{\vec{k}}$, and the other entries obtained as $A_i = (1 \leftrightarrow i) \cdot A_1$.
For instance, for $P=Z$ and $\vec{k}=[0,0,0, n-1]$ we get 
a vector of observables with entries
\begin{equation}\label{ex:equiv_bn_to_vect}
    A_j = Z_j.
\end{equation}
For $P=Z$ and $\vec{k}=(0,0,1, n-2)$, we would rather get
\begin{equation}
    \;A_j = Z_j \sum_{k\neq j} Z_k.
\end{equation}

\para{Case III: $(V=\hilbop, W=\mathbb{R}^{n \times n})$}
Here, the equivariant maps are identified as matrices of observables $\mathbf{A} = \{ A_{j_1, j_2} \}$ that satisfy equivariance constraints~\eqref{eq:equiv_operators_tensor}. Now, each of the $n^2$ observables per matrix can be obtained from two entries, either $A_{1,1}$ for the diagonal elements $A_{i,i}$ or $A_{1,2}$ for the off-diagonal ones.
The diagonal elements are subject to exactly the same constraints as before (case II) giving $4\times {\rm Te}(n)$ independent maps (diagonal matrices of observables).
For the off-diagonal ones, the equivalence classes are labeled by a triplet $(P_1, P_2, \vec{k})$ consisting of two Pauli operators $P_1, P_2$ and a valid composition $\vec{k}\vdash (n-2)$ on the $n-2$ remaining qubits; we have $4^2 \times {\rm Te}(n-1)$ of such equivalence classes.
For these, $A_{1,2} = P_1 \otimes P_2 \otimes S_{\vec{k}}$ and the other entries $A_{i, j\neq i}$ are obtained through $(1 \leftrightarrow i)\cdot(2\leftrightarrow j) \cdot A_{1,2}$.

As an example of an off diagonal matrix observables, taking $P_1 = P_2=Z$ and $\vec{k}=[0,0,0, n-2]$ we get a matrix of off-diagonal observables
\begin{equation}\label{ex:equiv_bn_to_graph}
    A_{i,j\neq i}=Z_i Z_j.
\end{equation}
For $P_1=Z$, $P_2=I$ and $\vec{k}=[0,0,1, n-3]$ we rather get
\begin{equation}\label{ex:equiv_bn_to_graph_2}
    A_{i,j\neq i}= Z_i \sum_{k \notin \{i,j\}} Z_k
\end{equation}

\para{Case III bis: $(V=\mathbb{R}^{n \times n}, W=\hilbop)$}
Exchanging $V$ and $W$, we now have maps from graphs to observables, with $\dket{P_{\vec{o}}}\dbra{j_1,j_2}$ mapping an input matrix $x$ to the observable $x_{j_1,j_2} P_{\vec{o}}$.
Then, for instance, Eq.~\eqref{ex:equiv_bn_to_graph} becomes
\begin{equation}\label{ex:equiv_graph_to_bn}
    x \mapsto H_{ZZ}(x) := \sum_{i \neq j} x_{i,j} Z_i Z_j
\end{equation}
that maps a graph to a weighted sum of Ising terms $ZZ$.
We highlight that Eq.~\eqref{ex:equiv_graph_to_bn}, together with $S_X$~\eqref{ex:symm_operators}, are the generators of the parameterized circuits used in the quantum approximate optimization algorithm (QAOA)~\cite{farhi2014quantum}.
For Eq.~\eqref{ex:equiv_bn_to_graph_2}, we rather get the map 
\begin{equation}\label{ex:equiv_graph_to_bn_2}
    x \mapsto \sum_{i \neq j} Z_i Z_j \left( \sum_{k \notin{\set{i,j}}} x_{i,k} + \sum_{k \notin{\set{i,j}} } x_{j,k} \right).
\end{equation}

\para{General case: $(V=\mathbb{R}^{n^k}, W=\hilbop)$} Cases I, II and III ($k=0$, $1$ and $2$) can be generalized to arbitrary $k$. For that we wish to find the equivalence classes of the set $\set{\dket{j_1, \hdots j_k}\dbra{P_{o_1} \hdots P_{o_n}}}$.
First, according to Sec.~\ref{sec:ex_classical}, we know that we can group the classical indices in terms of partitions of the $k$ indices. Each partition corresponds to the entries of the order-$k$ tensor of observables that are needed to identify all $n^k$ observables. 
Then, a partition of $k$ into $k'$ non-empty subsets fixes $k'$ qubits entails a conjugacy class indexed by all possible Pauli strings on the first $k'$ qubits (with $4^{k'}$ of them), together with a valid composition over the $n-k'$ remaining qubits (with ${\rm Te}(n-k'-1)$ of them).
Finally, recalling that the number of partitions with $k'$ non-empty subsets are the Stirling number of the second kind ${k \brace k'}$, and summing over all $k'$ we retrieve the dimension reported in Table~\ref{tab:equiv_maps} (G2, third line). These maps could be used, for instance, when dealing with hypergraphs (the case $k>2$).

\subsubsection{Specialization}\label{sec:special}
In a  classical ML setting, for a fixed choice of input and output domain, all the equivariant maps would have roughly the same implementation cost.
However, in QML the cost of implementing a given map may vary significantly depending on its specificity.
As such, it is often desirable to restrict the space of maps to the ones that can be implemented in reasonable effort.
To do so, we specialize the previous results to incorporate natural restrictions, both for measurements and circuit constructions, by considering subspaces of observables.

Measuring expectation values of non-commuting observables requires measurements in various bases (either deterministically or randomly chosen).
To reduce the sampling overhead, it can thus be interesting to only consider observables diagonal in a given basis.
For instance, all observables belonging to $\hilbop_{\{I,Z\}}$, defined as the subspace of observables spanned by Pauli strings composed uniquely of $I$ and $Z$, can be simultaneously measured in the computational basis. 
Additionally, when implementing gates through exponentiation of generators, one typically would like to reduce the locality of the generators. Then, one would consider $\mc{B}_{(l)}$ defined as the subspace spanned by Pauli strings with a small number $l$ (the \emph{locality}) of non-identity Pauli operators composing the strings: In our notation, for a Pauli string $P_{\vec{o}}$ on $n$ qubits, $l=n - k_i(\vec{o})$.

Notably, the restricted spaces $\hilbop_{\{I,Z\}}$ and $\hilbop_{(l)}$ are closed under the action of permutations and admit as a basis a subset of the Pauli strings. Hence the methodology of Sec.~\ref{sec:twirl} directly applies. The most natural examples of such restrictions are reported in Table~\ref{tab:equiv_maps} where we provide the dimensions of the spaces of equivariant maps obtained. This includes equivariant and invariant maps that can be used for measurements on the computational basis (M1-M3, fourth to seventh line) and symmetric and equivariant observables at most two-local, with $l\leq 2$ (E1-E2, eight to last line).

\subsection{Non-linear equivariant maps}\label{sec:equiv_nonlinear_maps}
Having characterized all relevant equivariant linear maps, it remains to characterize the non-linear ones. Compared to the previous such characterization is much more straightforward.

Classically, nonlinearities respecting equivariance are achieved through \emph{element-wise} functions~\cite{bronstein2021geometric}, whereby a function $f:\mathbb{R}\mapsto \mathbb{R}$ is applied to each entry of a tensor $T\in \mathbb{R}^{n^k}$. 
The analogous in GQML are rather \emph{matrix functions} that can be shown to be generically equivariant: Consider an observable $A$ with eigendecomposition $A=\sum_i \lambda_i \ket{\lambda_i}\bra{\lambda_i}$, we see that any $f:\mathbb{R}\mapsto \mathbb{R}$, extended as a matrix function $f_{\rm mat}:\mc{B} \mapsto \mc{B}$ through $f_{\rm mat}(A) = \sum f(\lambda_i) \ket{\lambda_i}\bra{\lambda_i}$, satisfies  
\begin{equation}
f_{\rm mat}(R_q(\sigma) A R_q(\sigma)^{\dag}) = R_q(\sigma) f_{\rm mat}(A) R_q(\sigma)^{\dag}    
\end{equation}
That is, any matrix function is equivariant (irrespective on the underlying group of symmetries) as per Definition~\ref{def:equiv}. 
While quantum signal processing techniques~\cite{martyn2021grand} could enable realization of other non-linearity, in what follows we limit ourselves to exponentiation map $A \mapsto \exp(-i A)$ that is amenable to unitary implementation.

\section{GQML models for graphs: composition and improvements}\label{sec:composition}
Equipped with the previous characterization of their constituents, we are now in measure to recover, categorize and generalize existing GQML models for graphs.
After detailing equivariant encodings in Sec.~\ref{sec:equivariant_encoding}, we present unified families of invariant and equivariant quantum models in Sec.~\ref{sec:symmetric_models}.
We connect these families to models found in the literature, showing how the latter can be generalized, and further discuss their integration with classical models.
Additional merits of these models are demonstrated in two dedicated numerical experiments.
In Sec.~\ref{sec:distinguishability}, it is evidenced how existing models can be made more expressive, and in Sec.~\ref{sec:pretraining}, benefits of simple classical pretraining strategies are probed.

\subsection{Equivariant encodings}\label{sec:equivariant_encoding}

Equivariance is preserved through composition, such that for a generator $H_{\theta} \in \aff(\mathbb{R}^{n \times n}, \mathcal{B})$ we have that the (non linear) map from graphs to unitaries $U(x) =\exp(-i H_{\theta}(x))$ is equivariant, since for any $\sigma$ and $\theta$
\begin{align}\label{eq:compo_1}
    \begin{split}
    U(\sigma \cdot x) &= \exp(-i R_q(\sigma) H_{\theta}(x) R_q(\sigma)^\dagger) \\
    &= R_q(\sigma) U(x) R_q(\sigma)^\dagger
    \end{split}
\end{align}
hold. Equivariance is further preserved through multiplication, in that for any equivariant $U(x)$ and $U'(x)$, we have 
\begin{equation}\label{eq:compo_2}
U(\sigma \cdot x) U'(\sigma \cdot x) = R_q(\sigma) U(x) U'(x) R_q(\sigma)^\dagger.
\end{equation}
These two properties allow us to define the equivariant encodings as follows.
\begin{definition}[Equivariant graph encoding]\label{def:equiv_enc}
Given a pure state $\ket{\psi_0}$ invariant under $S_n$ (with $R_{q}(\sigma) \cdot \ket{\psi_0} = \ket{\psi_0}$ for all  $\sigma \in S_n$) and a set of equivariant affine generator maps $H^{(l)}_{\theta}(x)\in \aff(\mathbb{R}^{n \times n},\mathcal{B})$~\eqref{def:equiv_affine_map}, 
the encoding $\rho_{\theta}(x)= \ket{\psi_\theta(x)}\bra{\psi_\theta(x)}$ is defined as
\begin{equation}\label{eq:equiv_enc}
\ket{\psi_\theta(x)}  = \prod_l e^{-iH^{(l)}_{\theta}(x)} \ket{\psi_0}.   
\end{equation}
From Eqs.~\eqref{eq:compo_1} and~\eqref{eq:compo_2}, one can 
verify equivariance~\eqref{eq:equiv} of these encodings: for all $\sigma$ and $\theta$,
\begin{equation*}
\rho_{\theta}(\sigma \cdot x) = R_{q}(\sigma) \rho_{\theta}(x) R_{q}(\sigma)^\dagger.
\end{equation*}
\end{definition}

A few comments are in order.
First, invariant pure states $\ket{\psi_0}$ are linear combinations of Dicke states~\cite{bartschi2019deterministic}, including $\ket{0}^{\otimes n}$, $\ket{+}^{\otimes n}$.
Second, the concept of equivariant encoding is not new, see for instance Refs.~\cite{mernyei2022equivariant}. In fact, equivariant graph encodings can be traced back to the beginning of variational quantum algorithms: the state prepared by the QAOA ansatz~\cite{farhi2014quantum} is precisely of the form Eq.~\eqref{eq:equiv_enc} with generators of the circuit given in Eqs.~\eqref{ex:symm_operators} and~\eqref{ex:equiv_graph_to_bn}
repeated $L$ times, such that 
\begin{equation}\label{def:qaoa}
    \ket{\psi^{QAOA}_\theta(x)}  = \left(\prod_{l=1}^L e^{-i \theta_{l} S_X} e^{-i \theta_{l+n} H_{ZZ}(x)} \right)\ket{+}^{\otimes n}   
\end{equation}
with $\theta\in \mathbb{R}^{2L}$.
Similarly, graph states~\cite{hein2006entanglement} are specific instances of equivariant graph encoding, that can be recovered with the additional generator $S_Z$ from Eq.~\eqref{ex:symm_operators} together with a particular choices of parameters.
Generalization of these known encodings, arise from the new (even when restricted to $2$-body generators) symmetric and equivariant observables from Table~\ref{tab:equiv_maps}.
Finally, while expressed here in the circuit model, we note that Definition~\ref{def:equiv_enc} readily extends to analog quantum computing with the invariant initial state $\ket{\psi_0}$ now evolved under a Hamiltonian of the form
\begin{equation}\label{eq:equiv_enc_analog}
H(t) = \sum_l H^{(l)}(x) f^{(l)}(t)   
\end{equation}
with again $H^{(l)} \in \aff(\mathbb{R}^{n \times n},\mathcal{B})$ but with parameters replaced by (possibly time-dependent) control fields $f^{(l)}(t)$.

\subsection{Invariant and equivariant models}\label{sec:symmetric_models}
The families of invariant and equivariant models are built on these equivariant encodings.
\begin{definition}[Invariant graph GQML models]\label{def:invariant_model_graph}
With $\rho_{\theta}(x)$ an equivariant graph encoding~\eqref{eq:equiv_enc}, and $O_{\alpha}\in \aff(\mathbb{R}^{n \times n}, \mathcal{B})$ an affine equivariant observable map~\eqref{def:equiv_affine_map}, an invariant quantum model $q_{\alpha, \theta}: \mathbb{R}^{n \times n} \mapsto \mathbb{R}$ is defined as
\begin{equation}\label{eq:all_qml_inv}
    q_{\alpha, \theta}(x) := {\rm Tr } \left[ O_{\alpha}(x)  \rho_{\theta}(x) \right].
\end{equation}
Given the equivariance of $\rho_{\theta}$ and $O_{\alpha}$,
one can verify the invariance~\eqref{eq:equiv} of these models: for all $\sigma$, $\theta$ and $\alpha$,
\begin{align*}
\begin{split}
    q_{\alpha, \theta}(\sigma \cdot x) &= {\rm Tr } \left[ R(\sigma) O_{\alpha}(x)  R(g)^{\dagger} R(g) \rho_{\theta}(x) R(g)^{\dagger} \right] \\
    &= q_{\alpha, \theta}(x).
\end{split}
\end{align*}
\end{definition}

To our knowledge, Definition~\ref{def:invariant_model_graph} encompasses all known invariant models found in the literature, including cases where the observable $O$ is symmetric~\cite{mills2019quantum,schatzki2024theoretical} (input-independent) or equivariant~\cite{szegedy2019qaoa} (input-dependent). 

Recalling from Sec.~\ref{sec:act_quantum},
that for a $k$-order tensor of observables $\mathbf{A}$, the corresponding tensor of expectation values is denoted as $\mathbf{A}[\rho]\in \mathbb{R}^{n^k}$ , families of equivariant models are defined as follows.

\begin{definition}[Equivariant graph GQML models]\label{def:equivariant_model_graph}
With $\rho_{\theta}(x)$ an equivariant graph encoding~\eqref{eq:equiv_enc}, and $\mathbf{A}_{\alpha}$ an order-$k$ equivariant tensor of observables (Definition~\ref{def:equiv_operators_tensor}), an equivariant quantum model $q_{\alpha, \theta}: \mathbb{R}^{n \times n} \mapsto \mathbb{R}^{n^k}$ is defined as
\begin{equation}\label{eq:all_qml_equiv}
    q_{\alpha, \theta}(x) := \mathbf{A}_{\alpha} \left[\rho_{\theta}(x) \right].
\end{equation}
Given the equivariance of $\rho_{\theta}$ and $O_{\alpha}$,
one can verify the equivariance~\eqref{eq:equiv} of these models: for all $\sigma$, $\theta$ and $\alpha$:
\begin{equation*}
    q_{\alpha, \theta}(\sigma \cdot x) = \sigma \cdot \mathbf{A}_{\alpha} \left[\rho_{\theta}(x) \right].
\end{equation*}
\end{definition}
Again, such a definition captures, to our knowledge, all known models found in the literature. These have been limited to $k=1$~\cite{mernyei2022equivariant,skolik2023equivariant,albrecht2023quantum} and sometimes $k=2$~\cite{thabet2023enhancing}, with only the simplest choice of vectors or matrix of observables utilized, such as Eqs.~\eqref{ex:equiv_bn_to_vect} and~\eqref{ex:equiv_bn_to_graph}.

Overall, through Definitions~\ref{def:invariant_model_graph} and~\ref{def:equivariant_model_graph} one can compose rich families of graph QML models: These unify known models, through our definitions of affine equivariant maps~\eqref{def:equiv_affine_map}, while generalizing them through the complete characterization of these maps in Table~\ref{tab:equiv_maps}. 
Admittedly, these maps represent more that can be assessed in current quantum computing platforms. 
Still, even when restricting ourselves to subset of such maps that admit low overhead (second part of Table~\ref{tab:equiv_maps}), these already offer opportunities to improve on existing models. This is explored in numerical experiments in Sec.~\ref{sec:distinguishability}.
Before that, we detail how these models can be interfaced  with classical ones.

\subsection{Interfacing equivariant quantum and classical models}\label{sec:interfacing}
As discussed, it seems preferable to employ QML models as a specialized part of an ML framework rather than requiring them to solve end-to-end problems.
The families of invariant and equivariant quantum models lend themselves to simple integration with (classical) GML models through \emph{data extension}. 
As depicted in Fig.~\ref{fig:summary}(d), one would task the QML models to output invariant or equivariant data that can be appended to the input data as additional features and then processed classically. In fact, such an extension is common practice in overcoming the limitations of GNNs~\cite{bouritsas2022improving, puny2023equivariant}. 

Consider an input graph $x \in \mathbb{R}^{n \times n}$ and an 
equivariant model (Definition~\ref{def:equivariant_model_graph}) for $k=2$ that has output matrix $q(x)\in \mathbb{R}^{n \times n}$ satisfying $q(\sigma \cdot x)= \sigma \cdot q(x)$.
One would then extend $x$ to $\tilde{x} \in \mathbb{R}^{n \times n \times 2}$ by appending $q(x)$ to the additional dimension as 
\begin{equation*}
    [\tilde{x}(x)]_{i,j,1}=x_{i,j}, \,\textrm{and}\; [\tilde{x}(x)]_{i,j,2}=[q(x)]_{i,j}.
\end{equation*}
Upon input transformation $\sigma \cdot x$, the extension reads
\begin{equation*}
    [\tilde{x}(\sigma \cdot x)]_{i,j,1}=[\sigma \cdot x]_{i,j}, \,\textrm{and}\; [\tilde{x}(\sigma \cdot x)]_{i,j,2}=[\sigma \cdot q(x)]_{i,j}.
\end{equation*}
That is, $\tilde{x}(\sigma \cdot x) = \sigma \cdot \tilde{x}(x)$\footnote{Recall that for a featured graphs $\tilde{x} \in \mathbb{R}^{n\times n \times F}$, with $F$ accounting for the features, $\sigma$ acts only on the first two dimensions
.} and thus is equivariant. 
While discussed for $k=2$, such extension works for arbitrary $k$. For instance, for $k=1$ only diagonal entries of the features would be filled with the the vector model output $q(x) \in \mathbb{R}^n$ such that $[\tilde{x}(x)]_{i,i,2}=q(x)_{i}$ and $[\tilde{x}(x)]_{i,j \neq i,2}=0$. 
For $k=0$ one could rather fill the additional entries with the scalar model output $q(x) \in \mathbb{R}$ such that $[\tilde{x}(x)]_{i,j,2}=q(x)$.

\subsection{Increasing distinguishability of existing models}\label{sec:distinguishability}
Existing invariant graph QML models have been obtained through the use of one measurement observable -- be it input-independent~\cite{mills2019quantum,schatzki2024theoretical} or dependent~\cite{szegedy2019qaoa} -- diagonal in the computational basis.
Similarly, equivariant models have involved one equivariant vector of observables in the computational basis~\cite{mernyei2022equivariant,skolik2023equivariant,albrecht2023quantum}.
Notably, in all these cases, only a \emph{single} observable, or a \emph{single} vector, was employed. Our characterization in Table~\ref{tab:equiv_maps} rather indicates that with exactly the same measurement basis (or equivalently, the same measurement dataset) one could have estimated many more linearly independent observables: the dimension of M1, M2 and M3. Hence, as depicted in Fig.~\ref{fig:summary}~(d), one could have estimated many more expectation values complying with equivariance principles. That is, it seems possible to render the underlying models more expressive, by incorporating these additional measurements, at virtually no cost (simple postprocessing).

To exemplify this in a minimal example, we revisit the task addressed in Ref~\cite{szegedy2019qaoa} that aimed at distinguishing \emph{non-isomorphic} graphs through the use of QAOA~\cite{farhi2014quantum}. (See also Refs.~\cite{mills2019quantum} for related approaches.) In our terminology and notations, the model $q: \mathbb{R}^{n \times n} \mapsto \mathbb{R}$ employed in Ref~\cite{szegedy2019qaoa} consists in the equivariant encoding of Eq.~\eqref{def:qaoa} and the equivariant observable of Eq.~\eqref{ex:equiv_bn_to_graph}.
According to Definition~\ref{def:invariant_model_graph}, the resulting model is invariant. Thus for two isomorphic graphs $x \cong x'$, such that $x'= \sigma \cdot x$ for some $\sigma \in S_n$, we are guaranteed that $q(x')=q(x)$. In other words, the model is a \emph{graph invariant} that can be used for the purpose of distinguishing non isomorphic graphs: if $q(x)\neq q(x')$ then we are guaranteed that $x \ncong x'$, although the converse is not true.

\begin{table}[t]
\centering
\begin{tabular}{l c c c c}
\hline
\textbf{Family}
& \textbf{Nodes} & \textbf{\# graphs} & \textbf{One obs.} & \textbf{Several obs.} \\
\hline\hline
Miyazaki & 20& 2& 3 & \textbf{1} \\
Praust & 20&2& 3 & \textbf{1}  \\
3-regular  &16&4060&4 &  \textbf{1}\\
3-regular  &18&41301& 4 & \textbf{1} \\
SRG 26,10,3,4   &26&10& 2 & \textbf{1} \\
\hline
\end{tabular}
\caption{Distinguishing non-isomorphic graphs. For different families of non-isomorphic graphs (with characteristics provided in the first three columns), we report the minimum depth $L$ of the QAOA encodings~\eqref{def:qaoa} enabling distinguishability of all the pairs of graphs (last two columns). These depths are reported for a model consisting of a single observable~\eqref{ex:equiv_graph_to_bn} as per Ref.~\cite{szegedy2019qaoa} (fourth column), and when supplemented with additional observables, diagonal in the same basis, that have been identified to yield invariant models (last column).
For the latter, the circuits are always strictly smaller. In all cases the measurements are not subjected to shot noise.}
\label{tab:qaoa_expressivity}
\end{table}

Ref.~\cite{szegedy2019qaoa} assesses the minimum number of layers $L$ of the encoding~\eqref{def:qaoa} that are necessary to distinguish all pairs of non-isomorphic graphs belonging to a given family of graphs.
The results are reported in Table~\ref{tab:qaoa_expressivity} for different families of graphs (first column) corresponding to different number of nodes (second column) and cardinality (third column). 
In line with Ref.~\cite{szegedy2019qaoa}, it is found that for the single observable of Eq.~\eqref{ex:equiv_bn_to_graph}, a single layer is never sufficient to distinguish all pairs, but a number of layers $L\in \{2, 3, 4\}$ is necessary (fourth column).
Rather, when keeping the same circuit structure and measurement basis, but incorporating all additional compatible observables (from Table~\ref{tab:equiv_maps}, M1-M2) we still retain the graph invariant nature of the algorithm.
Notably, with such an extension we are able to reduce the  circuit sizes to their bare minimum $L=1$ (fifth column).

It should be stressed that these results do not mean that the extended model solves the graph isomorphism problem since only a very small subset of graphs (albeit typically hard to distinguish) were considered, and as it is assumed that expectation values are obtained up to numerical accuracy (not accounting for shot noise).
However, this shows that the extended model was made strictly more expressive than the original one, and that, at no cost (except for additional classical computations of expectation values).

\begin{figure*}[t!]
\includegraphics[width=0.95\textwidth]{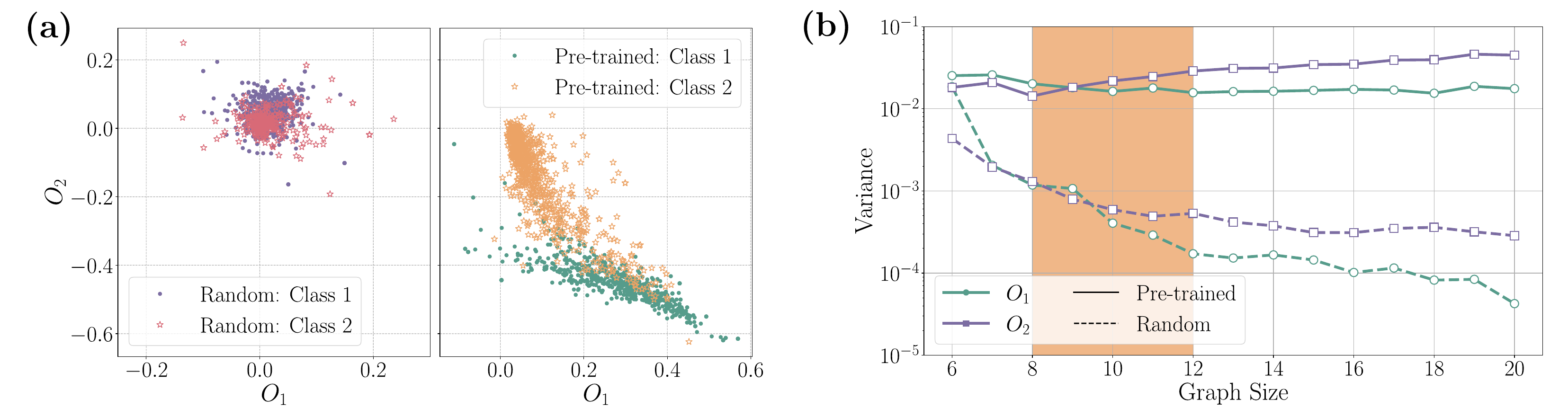}
\vspace{-10pt}\caption{Classical pre-training. The invariant QML model considered has encoding circuit defined through Eq.~\eqref{def:qaoa_extended} and a $2$-dimensional output given by the expectation values of $2$ observables $O_1$ and $O_2$~\eqref{eq:obs_model}. \textbf{(a)} For larger graphs ($n=14$ to $20$ nodes) than used for training ($n=8$ to $12$ nodes), outputs of the classically pre-trained model (right panel) are compared to output of the model initialized with random parameters (left panel). Each point correspond to one graph and has coordinates $(x,y)$ being the observables expectation values. \textbf{(b)} For a more detailed comparison, the dispersion (quantified by the variance) of the output of the models are reported at varied system size and for the two observables $O_1$ (left panel) and $O_2$ (right panel).
}
\label{fig:pre_training}
\end{figure*}

\subsection{Pre-training}\label{sec:pretraining}
A notable feature of the parameterized maps found in Table~\ref{tab:equiv_maps} is that they do not depend on the input size. For instance, the equivariant map $x \mapsto \sum_{ij} x_{ij} Z_iZ_j$~\eqref{ex:equiv_graph_to_bn}, is equally applicable to any graph $x$ independent of its size. In turn, the number of parameter of any of our graph models (Definitions~\ref{def:invariant_model_graph} and~\ref{def:equivariant_model_graph}) does not depend on the system size of the problems addressed.
This feature -- a direct consequence of the symmetries considered -- is in contrast to generic QML models.

In turn, the many training strategies~\cite{streif2020training,wurtz2021fixed,sack2021quantum,sauvage2021flip,shaydulin2023parameter,goh2025lie,pelofske2026evaluating} that have been devised to facilitate the scaling of QAOA -- which as discussed is a particular instance of an equivariant encoding~\eqref{def:qaoa} -- can be ported to our models.
Among them, we focus on classical pre-training. Here, the model is trained on graph inputs small enough such that the training can be entirely performed classically. Then the obtained parameters can be used for inference on larger graphs, or as pre-initialized parameter values for further training, on a quantum device.
Appeal of this methodology lies in its simplicity, and is further motivated by phenomenon of concentration of good parameters in QAOA~\cite{akshay2021parameter,brandao2018fixed} whereby one expects that good parameters for a given graph (drawn from some distribution) are typically good parameters for other graphs (drawn from the same distribution). 

To assess the potential of this pre-training in a controlled setting, we consider a binary classification problem whereby graphs in the first class are \emph{tree graphs} while graphs from the second class are graphs generated via an Erdos-Renyi distribution with parameter $p=0.3$ that are \emph{cyclic} (ie that are not trees).
This has the advantage that graphs can be generated on demand at arbitrary system sizes, and with probability of generating twice the same graph decreasing exponentially with $n$. This allows us to easily generate data points for the training and testing to extensively probe the pre-training strategy.

The equivariant encoding~\eqref{eq:equiv_enc} employed consist of $L=10$ layers of a generalization of Eq.~\eqref{def:qaoa}. Each layer, indexed by $l \in [10]$ is generated by elements of Eqs.~\eqref{ex:symm_operators} and~\eqref{ex:equiv_graph_to_bn} (pertaining to the classes E1 and E2 of Table~\ref{tab:equiv_maps}):
\begin{align}\label{def:qaoa_extended}
    \begin{split}
    \ket{\psi_\theta(x)}  = \Big(\prod_{l=1}^L e^{-i \theta_{l} S_X} &e^{-i \theta_{l+n} S_Z} e^{-i \theta_{l+2n} S_X} \\
    &e^{-i \theta_{l+3n} H_{ZZ}(x)} \Big)\ket{+}^{\otimes n} ,  
    \end{split}
\end{align}
accounting for a total of $40$ circuit parameters.
Outputs of the model is specified in terms of the $2$ observables:
\begin{align}\label{eq:obs_model}
\begin{split}
    &O_1(x) = \sum_{i,j} \frac{x_{i,j}}{\| x\|_1} Z_iZ_j, \;\;
    O_2  = \sum_i \frac{1}{n}Z_i,
\end{split}
\end{align}
where $\| x \|_1 = \sum_{ij} |x_{i,j}|$\footnote{
$\|x\|_1$ and $n$ are used to rescale the operators to ensure $\mathcal{O}(1)$ operator norms. These quantities are global (ie invariant) properties of the graphs and as such do not change equivariance properties of the observable maps.} (twice the number of edges for simple unweighted graphs).
The $2$ features are used for linear classification through a support vector machine model, introducing $2$ more parameters describing the hyperplane separating the $2$ classes. Training is performed over $40$ graphs of size $n \in [8,12] $. For this relatively simple problem the trained model achieves accuracy of $85\%$.

To assess the pre-training, we apply the pretrained model to both graphs of smaller sizes $ n \in [6, 7]$) and larger size ($n \in [13, 20]$). Results are reported in Fig.~\ref{fig:pre_training}. 
For a qualitative picture, Fig.~\ref{fig:pre_training}(a) displays output of the pre-trained model compared to the output of a similar model but with a set of parameters  drawn at random.  
As can be seen, the pre-trained model produces expectation values that allow to accurately distinguish between the two classes (green circles and orange stars) with relatively clear separations.
In comparison, values of the randomly initialized models concentrate around zero and display strong overlaps between the two classes (purple dots and red stars) impeding differentiating the classes.

To more precisely analyze the scalings at play, we report variances in the expectation values of the two observables~\eqref{eq:obs_model} (left and right panels) at varied system sizes, $n \in [6, 20]$, in Fig.~\ref{fig:pre_training}(b).
For each $n$, $50$ graphs for each classes are generated and based on these, variances in the expectation values are computed.
For the randomly initialized model (light green curve), we see  phenomenon of concentration whereby variances appear to decay exponentially with the system size.
In the case of pre-trained models (purple) we see that the variances remains steady, and with magnitudes consistently larger, with up to more than $2$ orders of magnitude for the largest sizes studied. 
Overall, we see that on this problem,  pre-training significantly reduces issues of concentration. This simple strategy can readily be extended to more elaborate strategies, such as gradually increasing the size of the graphs seen during the training, or the number of layers.

\section{Summary and Outlook}\label{sec:conclusion}

In this work, we detailed a toolbox for the construction of GQML model on graphs, both invariant (Definition~\ref{def:invariant_model_graph}) and equivariant (Definition~\ref{def:equivariant_model_graph}), and demonstrated its appealing features: known models are unified and extended, integration with classical models is natural, and simple but powerful pre-training is readily performed.

Symmetrization (Sec.~\ref{sec:twirl}) was central to the design of the toolbox and allowed us to identify all relevant equivariant linear (and thus affine) maps (Sec.~\ref{sec:equiv_linear_maps}) that were reported in Table~\ref{tab:equiv_maps}.
Including non-linear equivariant maps (Sec.~\ref{sec:equiv_nonlinear_maps}) and focusing on those that can be implemented straightforwardly, we identified general families of symmetric models (Sec.~\ref{sec:symmetric_models}). As was discussed, these contain and generalize known instances of symmetric models of graphs that we are aware of.
Integration with classical models was further detailed (Sec.~\ref{sec:interfacing}).
In particular, we advanced data extension (through additional features generated by the quantum models) as a natural way to integrate quantum and classical models complying with symmetric requirements.

In addition, the benefits of the toolbox were demonstrated in dedicated numerical experiments. It was shown that known models could be made more expressive without incurring additional  overhead (Sec.~\ref{sec:distinguishability}). Furthermore, the use of a classical pre-training strategy was studied, which, despite its simplicity, was shown to substantially mitigate concentration issues (Sec.~\ref{sec:pretraining}).

Going further, we hope this work will encourage similar studies for other families of problems, and related group of symmetries. 
While the use of symmetrization was found to be particularly appropriate for the symmetric group $S_n$, 
it remains to be seen how practical it will be for other group of symmetries.
We further note that equivariant models (with tensor outputs rather than scalar ones) have so far been limited to $S_n$.
Our treatment of equivariance through the tensor of observables~\eqref{eq:equiv_operators_tensor} -- that relates to representation tensors found in classical GML~\cite{cohen2016group} -- opens the path to achieving equivariance more systematically. As was discussed, this will be pivotal to rich integration of quantum and classical models.

Taken altogether, these studies and results provide important steps towards principled design and application of QML models for graphs. 
Still, some specific design choices were made along the way and our numerical experiments limited;
There remain many opportunities for further studies of the underlying models and towards their applications to end-to-end problems. 

On the practical side, utility of QML models at sizes exceeding classical computations remains to be ascertained. This includes, as a pre-requisite, demonstration of non-vanishing 
outputs of the quantum models at larger scales than probed here. This will challenge viability of pre-training, or ad-hoc initialization, strategies as envisioned here and elsewhere. In addition, this will require training and application to datasets of interest for the community~\cite{Morris+2020}. 

On a more theoretical side, understanding what the proposed families of models can compute will be key to further advances.
Progresses in the field of GML, in particular for graphs, have been initiated by the identification of limitations in standard GNNs, and have motivated the design of novel architectures to palliate these limitations~\cite{bouritsas2022improving,xu2018powerful,morris2019weisfeiler,puny2023equivariant}.
Similarly, identifying limitations of GQML models would guide the design of novel architectures.
While we limited ourselves to encoding $n$-node graphs into $n$-qubit systems, other encoding quantum spaces (or equivalently, other actions or representations of the $S_n$) are of interest. These include the use of $n^2$ qubits to encode $n$-node graphs, or the use of multiple copies.
In particular, the latter can be the source of exponential separation in sampling complexity~\cite{aharonov2022quantum,huang2022quantum}.
Similarly, we restricted ourselves to non-linearity by means of exponentiation (yielding unitaries) but realization of other kind of non-linearity through quantum singular value transforms is of interest~\cite{martyn2021grand}. 
All these can lead to more powerful quantum models at the cost of increased implementation burden.
With the ongoing improvements of quantum computing technology, the study of richer models will become more and more relevant.

\section{Acknowledgments}
We thank Colin Krawchuk and Mateusz Kupper for their feedback on the manuscript. ML acknowledges support by the Laboratory Directed Research and Development (LDRD) program of LANL under project number 20230049DR and 20260043DR, by the LANL's ASC Beyond Moore’s Law project.

\clearpage
\bibliography{biblio.bib}

@inproceedings{cohen2016group,
  title={Group equivariant convolutional networks},
  author={Cohen, Taco and Welling, Max},
  booktitle={International conference on machine learning},
  pages={2990--2999},
  year={2016},
  organization={PMLR},
  url = 	 {https://proceedings.mlr.press/v48/cohenc16.html}
}

@article{bronstein2021geometric,
  title={Geometric deep learning: Grids, groups, graphs, geodesics, and gauges},
  author={Bronstein, Michael M and Bruna, Joan and Cohen, Taco and Veli{\v{c}}kovi{\'c}, Petar},
  journal={arXiv preprint arXiv:2104.13478},
  year={2021},
url={https://doi.org/10.48550/arXiv.2104.13478}
}

@inproceedings{kondor2018generalization,
  title={On the generalization of equivariance and convolution in neural networks to the action of compact groups},
  author={Kondor, Risi and Trivedi, Shubhendu},
  booktitle={International conference on machine learning},
  pages={2747--2755},
  year={2018},
  organization={PMLR},
  url = 	 {https://proceedings.mlr.press/v80/kondor18a.html}
}

@article{xu2018powerful,
  title={How powerful are graph neural networks?},
  author={Xu, Keyulu and Hu, Weihua and Leskovec, Jure and Jegelka, Stefanie},
  journal={arXiv preprint arXiv:1810.00826},
  year={2018},
  url={https://openreview.net/forum?id=ryGs6iA5Km},
}

@inproceedings{maron2018invariant,
title={Invariant and Equivariant Graph Networks},
author={Haggai Maron and Heli Ben-Hamu and Nadav Shamir and Yaron Lipman},
booktitle={International Conference on Learning Representations},
year={2019},
url={https://openreview.net/forum?id=Syx72jC9tm}
}

@inproceedings{morris2019weisfeiler,
  title={Weisfeiler and leman go neural: Higher-order graph neural networks},
  author={Morris, Christopher and Ritzert, Martin and Fey, Matthias and Hamilton, William L and Lenssen, Jan Eric and Rattan, Gaurav and Grohe, Martin},
  booktitle={Proceedings of the AAAI conference on artificial intelligence},
  number={01},
  pages={4602--4609},
  year={2019},
  url={
https://doi.org/10.48550/arXiv.1810.02244}
}

@article{bouritsas2022improving,
  title={Improving graph neural network expressivity via subgraph isomorphism counting},
  author={Bouritsas, Giorgos and Frasca, Fabrizio and Zafeiriou, Stefanos and Bronstein, Michael M},
  journal={IEEE Transactions on Pattern Analysis and Machine Intelligence},
  volume={45},
  number={1},
  pages={657--668},
  year={2022},
  publisher={IEEE},
  url={https://doi.org/10.1109/TPAMI.2022.3154319}
}

@inproceedings{puny2023equivariant,
  title={Equivariant polynomials for graph neural networks},
  author={Puny, Omri and Lim, Derek and Kiani, Bobak and Maron, Haggai and Lipman, Yaron},
  booktitle={International Conference on Machine Learning},
  pages={28191--28222},
  year={2023},
  organization={PMLR},
  url = 	 {https://proceedings.mlr.press/v202/puny23a.html},
}

@article{zhu2020beyond,
  title={Beyond homophily in graph neural networks: Current limitations and effective designs},
  author={Zhu, Jiong and Yan, Yujun and Zhao, Lingxiao and Heimann, Mark and Akoglu, Leman and Koutra, Danai},
  journal={Advances in neural information processing systems},
  volume={33},
  pages={7793--7804},
  year={2020},
  url={https://doi.org/10.48550/arXiv.2006.11468}
}

@article{alon2020bottleneck,
  title={On the bottleneck of graph neural networks and its practical implications},
  author={Alon, Uri and Yahav, Eran},
  journal={arXiv preprint arXiv:2006.05205},
  year={2020},
  url={https://doi.org/10.48550/arXiv.2006.05205}
}

@inproceedings{cohen2017steerable,
  title={Steerable CNNs},
  author={Cohen, Taco S and Welling, Max},
  booktitle={International Conference on Learning Representations},
  year={2017},
  url={https://openreview.net/forum?id=rJQKYt5ll}
}

@inproceedings{finzi2021practical,
  title={A practical method for constructing equivariant multilayer perceptrons for arbitrary matrix groups},
  author={Finzi, Marc and Welling, Max and Wilson, Andrew Gordon},
  booktitle={International conference on machine learning},
  pages={3318--3328},
  year={2021},
  organization={PMLR},
  url = 	 {https://proceedings.mlr.press/v139/finzi21a.html}
}

@article{krizhevsky2012imagenet,
  title={Imagenet classification with deep convolutional neural networks},
  author={Krizhevsky, Alex and Sutskever, Ilya and Hinton, Geoffrey E},
  journal={Advances in neural information processing systems},
  volume={25},
  year={2012},
  url={https://doi.org/10.1145/306538}
}

@article{cerezo2022challenges,
  title={Challenges and opportunities in quantum machine learning},
  author={Cerezo, Marco and Verdon, Guillaume and Huang, Hsin-Yuan and Cincio, Lukasz and Coles, Patrick J},
  journal={Nature computational science},
  volume={2},
  number={9},
  pages={567--576},
  year={2022},
  publisher={Nature Publishing Group US New York},
  url={https://doi.org/10.1038/s43588-022-00311-3}
}

@article{mills2019quantum,
  title={Quantum invariants and the graph isomorphism problem},
  author={Mills, PW and Rundle, RP and Samson, JH and Devitt, Simon J and Tilma, Todd and Dwyer, VM and Everitt, Mark J},
  journal={Physical Review A},
  volume={100},
  number={5},
  pages={052317},
  year={2019},
  publisher={APS},
url={https://doi.org/10.1103/PhysRevA.100.052317}
}

@article{schatzki2024theoretical,
  title={Theoretical guarantees for permutation-equivariant quantum neural networks},
  author={Schatzki, Louis and Larocca, Martin and Nguyen, Quynh T and Sauvage, Frederic and Cerezo, Marco},
  journal={npj Quantum Information},
  volume={10},
  number={1},
  pages={12},
  year={2024},
  publisher={Nature Publishing Group UK London},
  url={https://doi.org/10.1038/s41534-024-00804-1}
}

@article{szegedy2019qaoa,
  title={What do QAOA energies reveal about graphs?},
  author={Szegedy, Mario},
  journal={arXiv preprint arXiv:1912.12277},
  year={2019},
  url={https://doi.org/10.48550/arXiv.1912.12277}
}

@inproceedings{mernyei2022equivariant,
  title={Equivariant quantum graph circuits},
  author={Mernyei, P{\'e}ter and Meichanetzidis, Konstantinos and Ceylan, Ismail Ilkan},
  booktitle={International Conference on Machine Learning},
  pages={15401--15420},
  year={2022},
  organization={PMLR},
  url={https://proceedings.mlr.press/v162/mernyei22a.html}
}

@article{skolik2023equivariant,
  title={Equivariant quantum circuits for learning on weighted graphs},
  author={Skolik, Andrea and Cattelan, Michele and Yarkoni, Sheir and B{\"a}ck, Thomas and Dunjko, Vedran},
  journal={npj Quantum Information},
  volume={9},
  number={1},
  pages={47},
  year={2023},
  publisher={Nature Publishing Group UK London},
  url={https://doi.org/10.1038/s41534-023-00710-y}
}

@article{albrecht2023quantum,
  title={Quantum feature maps for graph machine learning on a neutral atom quantum processor},
  author={Albrecht, Boris and Dalyac, Constantin and Leclerc, Lucas and Ortiz-Guti{\'e}rrez, Luis and Thabet, Slimane and D'Arcangelo, Mauro and Cline, Julia RK and Elfving, Vincent E and Lassabli{\`e}re, Lucas and Silv{\'e}rio, Henrique and others},
  journal={Physical Review A},
  volume={107},
  number={4},
  pages={042615},
  year={2023},
  publisher={APS},
  url={https://doi.org/10.1103/PhysRevA.107.042615}
}

@article{thabet2023enhancing,
  title={Enhancing Graph Neural Networks with Quantum Computed Encodings},
  author={Thabet, Slimane and Fouilland, Romain and Djellabi, Mehdi and Sokolov, Igor and Kasture, Sachin and Henry, Louis-Paul and Henriet, Lo{\"\i}c},
  journal={arXiv preprint arXiv:2310.20519},
  year={2023},
  url={https://doi.org/10.48550/arXiv.2310.20519}
}

@article{brandao2018fixed,
  title={For fixed control parameters the quantum approximate optimization algorithm's objective function value concentrates for typical instances},
  author={Brandao, Fernando GSL and Broughton, Michael and Farhi, Edward and Gutmann, Sam and Neven, Hartmut},
  journal={arXiv preprint arXiv:1812.04170},
  year={2018},
  url={https://doi.org/10.48550/arXiv.1812.04170}
}

@article{akshay2021parameter,
  title={Parameter concentrations in quantum approximate optimization},
  author={Akshay, Vishwanathan and Rabinovich, Daniil and Campos, Ernesto and Biamonte, Jacob},
  journal={Physical Review A},
  volume={104},
  number={1},
  pages={L010401},
  year={2021},
  publisher={APS},
url={https://doi.org/10.1103/PhysRevA.104.L010401}
}

@article{streif2020training,
  title={Training the quantum approximate optimization algorithm without access to a quantum processing unit},
  author={Streif, Michael and Leib, Martin},
  journal={Quantum Science \& Technology},
  volume={5},
  number={3},
  pages={034008},
  year={2020},
  publisher={IOP Publishing},
url={https://doi.org/10.1088/2058-9565/ab8c2b}
}

@article{wurtz2021fixed,
  title={Fixed-angle conjectures for the quantum approximate optimization algorithm on regular MaxCut graphs},
  author={Wurtz, Jonathan and Lykov, Danylo},
  journal={Physical Review A},
  volume={104},
  number={5},
  pages={052419},
  year={2021},
  publisher={APS},
  url={https://doi.org/10.1103/PhysRevA.104.052419}
}

@article{sack2021quantum,
  title={Quantum annealing initialization of the quantum approximate optimization algorithm},
  author={Sack, Stefan H and Serbyn, Maksym},
  journal={quantum},
  volume={5},
  pages={491},
  year={2021},
  publisher={Verein zur F{\"o}rderung des Open Access Publizierens in den Quantenwissenschaften},
url={https://doi.org/10.22331/q-2021-07-01-491}
}

@article{sauvage2021flip,
  title={Flip: A flexible initializer for arbitrarily-sized parametrized quantum circuits},
  author={Sauvage, Frederic and Sim, Sukin and Kunitsa, Alexander A and Simon, William A and Mauri, Marta and Perdomo-Ortiz, Alejandro},
  journal={arXiv preprint arXiv:2103.08572},
  year={2021},
  url={https://doi.org/10.48550/arXiv.2103.08572}
}

@article{shaydulin2023parameter,
  title={Parameter transfer for quantum approximate optimization of weighted maxcut},
  author={Shaydulin, Ruslan and Lotshaw, Phillip C and Larson, Jeffrey and Ostrowski, James and Humble, Travis S},
  journal={ACM Transactions on Quantum Computing},
  volume={4},
  number={3},
  pages={1--15},
  year={2023},
  publisher={ACM New York, NY},
  url={https://doi.org/10.1145/3584706}
}

@article{goh2025lie,
  title={Lie-algebraic classical simulations for quantum computing},
  author={Goh, Matthew L and Larocca, Martin and Cincio, Lukasz and Cerezo, Marco and Sauvage, Fr{\'e}d{\'e}ric},
  journal={Physical Review Research},
  volume={7},
  number={3},
  pages={033266},
  year={2025},
  publisher={APS},
  url={https://doi.org/10.1103/3y65-f5w6}
}

@article{pelofske2026evaluating,
  title={Evaluating the limits of Quantum Approximate Optimization Algorithm parameter transfer at high rounds on sparse Ising models with geometrically local cubic terms},
  author={Pelofske, Elijah and Rams, Marek M and B{\"a}rtschi, Andreas and Czarnik, Piotr and Braccia, Paolo and Cincio, Lukasz and Eidenbenz, Stephan},
  journal={Physical Review Research},
  volume={8},
  number={2},
  pages={023023},
  year={2026},
  url={https://doi.org/10.1103/p2lg-z4kn},
  publisher={APS}
}

@article{larocca2022group,
  title={Group-invariant quantum machine learning},
  author={Larocca, Mart{\'\i}n and Sauvage, Fr{\'e}d{\'e}ric and Sbahi, Faris M and Verdon, Guillaume and Coles, Patrick J and Cerezo, Marco},
  journal={PRX quantum},
  volume={3},
  number={3},
  pages={030341},
  year={2022},
  publisher={APS},
  url={https://doi.org/10.1103/PRXQuantum.3.030341}
}

@article{meyer2023exploiting,
  title={Exploiting symmetry in variational quantum machine learning},
  author={Meyer, Johannes Jakob and Mularski, Marian and Gil-Fuster, Elies and Mele, Antonio Anna and Arzani, Francesco and Wilms, Alissa and Eisert, Jens},
  journal={PRX quantum},
  volume={4},
  number={1},
  pages={010328},
  year={2023},
  publisher={APS},
url={https://doi.org/10.1103/PRXQuantum.4.010328}
}

@article{zheng2023speeding,
  title={Speeding up learning quantum states through group equivariant convolutional quantum ans{\"a}tze},
  author={Zheng, Han and Li, Zimu and Liu, Junyu and Strelchuk, Sergii and Kondor, Risi},
  journal={PRX quantum},
  volume={4},
  number={2},
  pages={020327},
  year={2023},
  publisher={APS},
  url={https://doi.org/10.1103/PRXQuantum.4.020327}
}

@article{nguyen2024theory,
  title={Theory for equivariant quantum neural networks},
  author={Nguyen, Quynh T and Schatzki, Louis and Braccia, Paolo and Ragone, Michael and Coles, Patrick J and Sauvage, Frederic and Larocca, Martin and Cerezo, Marco},
  journal={PRX Quantum},
  volume={5},
  number={2},
  pages={020328},
  year={2024},
  publisher={APS},
  url={https://doi.org/10.1103/PRXQuantum.5.020328}
}

@article{west2024provably,
  title={Provably trainable rotationally equivariant quantum machine learning},
  author={West, Maxwell T and Heredge, Jamie and Sevior, Martin and Usman, Muhammad},
  journal={PRX Quantum},
  volume={5},
  number={3},
  pages={030320},
  year={2024},
  publisher={APS},
  url={https://doi.org/10.1103/PRXQuantum.5.030320}
}

@article{le2025symmetry,
  title={Symmetry-invariant quantum machine learning force fields},
  author={Le, Isabel Nha Minh and Kiss, Oriel and Schuhmacher, Julian and Tavernelli, Ivano and Tacchino, Francesco},
  journal={New Journal of Physics},
  volume={27},
  number={2},
  pages={023015},
  year={2025},
  publisher={IOP Publishing},
  url={https://doi.org/10.1088/1367-2630/adad0c}
}

@book{fulton2013representation,
  title={Representation theory: a first course},
  author={Fulton, William and Harris, Joe},
  volume={129},
  year={2013},
  publisher={Springer Science \& Business Media},
  url={https://doi.org/10.1007/978-1-4612-0979-9}
}

@article{martyn2021grand,
  title={Grand unification of quantum algorithms},
  author={Martyn, John M and Rossi, Zane M and Tan, Andrew K and Chuang, Isaac L},
  journal={PRX quantum},
  volume={2},
  number={4},
  pages={040203},
  year={2021},
  publisher={APS},
  url={https://doi.org/10.1103/PRXQuantum.2.040203}
}

@article{farhi2014quantum,
  title={A quantum approximate optimization algorithm},
  author={Farhi, Edward and Goldstone, Jeffrey and Gutmann, Sam},
  journal={arXiv preprint arXiv:1411.4028},
  year={2014},
  url={https://doi.org/10.48550/arXiv.1411.4028}
}

@article{larocca2025barren,
  title={Barren plateaus in variational quantum computing},
  author={Larocca, Mart{\'\i}n and Thanasilp, Supanut and Wang, Samson and Sharma, Kunal and Biamonte, Jacob and Coles, Patrick J and Cincio, Lukasz and McClean, Jarrod R and Holmes, Zo{\"e} and Cerezo, M},
  journal={Nature Reviews Physics},
  pages={1--16},
  year={2025},
  publisher={Nature Publishing Group UK London},
  url={https://doi.org/10.1038/s42254-025-00813-9}
}

@article{toth2010permutationally,
  title={Permutationally invariant quantum tomography},
  author={T{\'o}th, G{\'e}za and Wieczorek, Witlef and Gross, David and Krischek, Roland and Schwemmer, Christian and Weinfurter, Harald},
  journal={Physical review letters},
  volume={105},
  number={25},
  pages={250403},
  year={2010},
  publisher={APS},
  url={https://doi.org/10.1103/PhysRevLett.105.250403}
}

@article{anschuetz2023efficient,
  title={Efficient classical algorithms for simulating symmetric quantum systems},
  author={Anschuetz, Eric R and Bauer, Andreas and Kiani, Bobak T and Lloyd, Seth},
  journal={Quantum},
  volume={7},
  pages={1189},
  year={2023},
  publisher={Verein zur F{\"o}rderung des Open Access Publizierens in den Quantenwissenschaften},
  url={https://doi.org/10.22331/q-2023-11-28-1189}
}

@article{sauvage2024classical,
  title={Classical shadows with symmetries},
  author={Sauvage, Frederic and Larocca, Martin},
  journal={arXiv preprint arXiv:2408.05279},
  year={2024},
url={https://doi.org/10.48550/arXiv.2408.05279}
}

@article{bacon2005quantum,
  title={The quantum Schur transform: I. efficient qudit circuits},
  author={Bacon, Dave and Chuang, Isaac L and Harrow, Aram W},
  journal={arXiv preprint quant-ph/0601001},
  year={2005},
  url={https://doi.org/10.48550/arXiv.quant-ph/0601001}
}

@article{biamonte2017quantum,
  title={Quantum machine learning},
  author={Biamonte, Jacob and Wittek, Peter and Pancotti, Nicola and Rebentrost, Patrick and Wiebe, Nathan and Lloyd, Seth},
  journal={Nature},
  volume={549},
  number={7671},
  pages={195--202},
  year={2017},
  publisher={Nature Publishing Group UK London},
  url={https://doi.org/10.1038/nature23474}
}

@article{benedetti2019parameterized,
  title={Parameterized quantum circuits as machine learning models},
  author={Benedetti, Marcello and Lloyd, Erika and Sack, Stefan and Fiorentini, Mattia},
  journal={Quantum science and technology},
  volume={4},
  number={4},
  pages={043001},
  year={2019},
  publisher={IOP Publishing},
  url={10.1088/2058-9565/ab4eb5}
}

@article{anschuetz2022quantum,
  title={Quantum variational algorithms are swamped with traps},
  author={Anschuetz, Eric R and Kiani, Bobak T},
  journal={Nature Communications},
  volume={13},
  number={1},
  pages={7760},
  year={2022},
  publisher={Nature Publishing Group UK London},
  url={https://doi.org/10.1038/s41467-022-35364-5}
}

@article{mcclean2018barren,
  title={Barren plateaus in quantum neural network training landscapes},
  author={McClean, Jarrod R and Boixo, Sergio and Smelyanskiy, Vadim N and Babbush, Ryan and Neven, Hartmut},
  journal={Nature communications},
  volume={9},
  number={1},
  pages={4812},
  year={2018},
  publisher={Nature Publishing Group UK London},
  url={https://doi.org/10.1038/s41467-018-07090-4}
}

@inproceedings{bartschi2019deterministic,
  title={Deterministic preparation of Dicke states},
  author={B{\"a}rtschi, Andreas and Eidenbenz, Stephan},
  booktitle={International Symposium on Fundamentals of Computation Theory},
  pages={126--139},
  year={2019},
  organization={Springer},
  url={https://doi.org/10.1007/978-3-030-25027-0_9}
}

@article{hein2006entanglement,
  title={Entanglement in graph states and its applications},
  author={Hein, Marc and D{\"u}r, Wolfgang and Eisert, Jens and Raussendorf, Robert and Nest, M and Briegel, H-J},
  journal={arXiv preprint quant-ph/0602096},
  year={2006},
  url={https://doi.org/10.48550/arXiv.quant-ph/0602096}
}

@inproceedings{Morris+2020,
    title={TUDataset: A collection of benchmark datasets for learning with graphs},
    author={Christopher Morris and Nils M. Kriege and Franka Bause and Kristian Kersting and Petra Mutzel and Marion Neumann},
    booktitle={ICML 2020 Workshop on Graph Representation Learning and Beyond (GRL+ 2020)},
    archivePrefix={arXiv},
    eprint={2007.08663},
    url={www.graphlearning.io},
    year={2020}
}

@article{aharonov2022quantum,
  title={Quantum algorithmic measurement},
  author={Aharonov, Dorit and Cotler, Jordan and Qi, Xiao-Liang},
  journal={Nature communications},
  volume={13},
  number={1},
  pages={887},
  year={2022},
  publisher={Nature Publishing Group UK London},
  url={https://doi.org/10.1038/s41467-021-27922-0}
}

@article{huang2022quantum,
  title={Quantum advantage in learning from experiments},
  author={Huang, Hsin-Yuan and Broughton, Michael and Cotler, Jordan and Chen, Sitan and Li, Jerry and Mohseni, Masoud and Neven, Hartmut and Babbush, Ryan and Kueng, Richard and Preskill, John and others},
  journal={Science},
  volume={376},
  number={6598},
  pages={1182--1186},
  year={2022},
  publisher={American Association for the Advancement of Science},
  url={https://doi.org/10.1126/science.abn7293}
}

@article{pearce2022connecting,
  title={Connecting permutation equivariant neural networks and partition diagrams},
  author={Pearce-Crump, Edward},
  journal={arXiv preprint arXiv:2212.08648},
  year={2022},
  url={https://doi.org/10.48550/arXiv.2212.08648}
}

\end{document}